\documentclass[12pt]{article}
\usepackage{amsmath}
\usepackage{graphics,graphicx,epsfig}
\usepackage{amssymb}
\usepackage{epstopdf}
\usepackage{sidecap}
\usepackage{appendix}
\DeclareFontFamily{OMX}{yhex}{}
\DeclareFontShape{OMX}{yhex}{m}{n}{<->yhcmex10}{}
\DeclareSymbolFont{yhlargesymbols}{OMX}{yhex}{m}{n}
\DeclareMathAccent{\wideparen}{\mathord}{yhlargesymbols}{"F3}

\begin{document}

\begin{titlepage}
\vfill
\begin{flushright}
UCPHAS-16-GR04
\end{flushright}

\vfill
\begin{center}
\baselineskip=16pt

{\Large\bf  Bianchi IX Cosmologies and the Golden Ratio}
\vskip 0.15in
\vskip 10.mm

{\bf  M S Bryant and D W Hobill}

\vskip 0.4cm
Department of Physics and Astronomy, University of Calgary, \\ Calgary, AB, T2N 2W1, Canada \\

\vskip 0.1 in Email: \texttt{mbryant@ucalgary.ca, hobill@ucalgary.ca}
\vspace{6pt}
\end{center}
\vskip 0.2in
\par
\begin{center}{\bf Abstract}
 \end{center}\begin{quote} 
Solutions to the Einstein equations for Bianchi IX cosmologies are examined through the use of Ellis-MacCallum-Wainwright (“expansion-normalized”) variables.  Using an
iterative map derived from the Einstein equations one can construct an infinite number of periodic solutions.  The simplest periodic solutions consist of 3-cycles.
It is shown that for 3-cycles the time series of the logarithms of the expansion-normalized spatial curvature components vs normalized time (which is runs backwards towards the initial singularity), generates a set of self-similar golden rectangles. In addition the golden ratio appears in other aspects of the same time series representation.

\vfill
\vskip 2.mm
\end{quote}
\hfill
\end{titlepage}

\section{Introduction}

Nonlinear dynamical systems can display a number of interesting phenomena such as deterministic chaos, self-similarity
(both continuous and discrete), pattern formation, critical behaviour, and self-organization. As a set of quasi-linear partial differential equations,
the Einstein equations are nonlinear in the metric and its first derivatives and thus form a nonlinear dynamical system.  Some of the better known nonlinear 
phenomena that occur in general relativity are 
the critical behaviour associated with gravitational collapse and black hole formation \cite{Gundlach} and the deterministic 
chaos occurring in the evolution of some anisotropic cosmologies \cite{Hobill}.  One feature of the critical phenomena in gravitational collapse is the 
existence of a discrete self-similarity that occurs at the onset of criticality.  While it is well known that a continuous self-similarity exists in the 
dynamics of Friedmann-Lemaitre-Robertson-Walker (FLRW) cosmologies, the existence of discrete self-similarities in cosmological solutions 
has not received a great deal of attention.  In this work we analyze one interesting example of a discrete self-similarity that arises in 
the dynamics of the vacuum
Bianchi IX cosmologies and show that it has properties that are related to the golden ratio, $\phi$. The number $\phi$ is interesting in
its own right, and it appears in disciplines as diverse as geometry, art, and biology. The fact that it also makes an appearance in general relativity
makes it all the more interesting (and curious).
 
One advantage of studying spatially homogeneous spacetimes is that the Einstein equations reduce to a set of nonlinear ordinary differential 
equations (ODEs) and this simplifies the subsequent analysis of the dynamics. Near the singularity it has been shown that important 
dynamical properties of the ODEs can be captured by algebraic and geometric iterative maps and this leads to further simplifications.     
One such map, the Gauss map, approximates the dynamics of the full Einstein 
equations for the Bianchi IX dynamics close to the cosmological singularity and it has both chaotic and periodic solutions. One periodic solution is related to $\phi$. 
The purpose of this presentation is to demonstrate that the 
golden ratio arises directly from the Einstein equations themselves without having to go through the dimensional reduction 
from differential equations to iterative maps.

In spite of their relative simplicity, the ODEs that govern the evolution of a homogeneous cosmological spacetime cannot be solved analytically 
except in very special cases and numerical
methods have been applied to studies of many homogeneous but anisotropic spacetimes. Given the diffeomorphism invariance of the Einstein
equations, different choices of time parameterization and dynamical variables have led to many controversies over how
those results might be interpreted (see \cite{Berger} for a review). 

\subsection{Bianchi Cosmologies: Bianchi IX and Kasner solutions} 

The Bianchi IX spacetimes have provided a fertile ground for studies of the time-dependent behaviour of the Einstein equations and the 
formation of spacetime singularities for close to
half a century.  Indeed attempts to understand the Bianchi IX cosmologies either for their own sake or as an 
approximation to the full inhomogeneous dynamics of the Einstein equations have been approached with a number of 
different formulations. Foremost among them are the Hamiltonian dynamical systems approach or the ``Mixmaster Universe" studies introduced by Misner 
\cite{Mis} and the Belinskii, Khalatnikov and Lifshitz (BKL) scale factor dynamics \cite{BKL,LL}
which have shed light on the complicated behaviour of non-standard cosmological spacetimes.  

Of particular interest has been the time-dependent behaviour of the Bianchi IX cosmologies close to the initial singularity.  In this case it can be expected
that curvature effects dominate over the  matter dynamics and one can study the pure vacuum equations. Furthermore BKL discovered that the Bianchi IX scale factors
could be approximated as a succession of transitions from one Bianchi I vacuum (Kasner) solution to another.    

The Bianchi I cosmology is the simplest homogeneous, anisotropic spacetime and its vacuum metric can be expressed in
terms of three explicitly time-dependent scale factors as \cite{Kas},
\begin{equation}
\label{KasnerMetric}ds^2=dt^2-t^{2p_1}dx^2-t^{2p_2}dy^2-t^{2p_3}dz^2
\end{equation}
 where $ p_1$, $p_2$ and $p_3$ are any three numbers that satisfy the conditions:
\begin{equation}
\label{eqnsforpis}p_1+p_2+p_3=1\quad\mbox{ and }\quad p_1^2+p_2^2+p_3^2=1\mbox{. }
\end{equation}
If two of the $p_i$s are the same, the only possible values are (0,0,1) or $ (-\frac{1}{3}$,$\frac{2}{3}$, $\frac{2}{3})$, but if the three values are all different, there is a continuum of possibilities.  They always lie within the ranges:
\begin{equation}
\label{piranges} -\frac{1}{3}\le p_1\le 0 \mbox{, }\quad 0\le p_2 \le\frac{2}{3}\quad \mbox{ and }\quad \frac{2}{3}\le p_3 \le1 \mbox{,}
\end{equation}
when the $p_i$s are arranged as $ p_1<p_2<p_3$.  The (0,0,1) case will be ignored in this paper, since it is an exception, resulting in a flat spacetime.

Based on the 
sign distribution (two positive and one negative) implied by (\ref{piranges}), the Kasner vacuum universe expands in two spatial dimensions and contracts in one.  
There is also a non-removable singularity at $t=0$.
Given the restrictions (\ref{eqnsforpis}) between the $p_i$s only a single parameter, $u$, is required to distinguish between different 
Kasner solutions. The $p_i$s in terms of the parameter $u$ are given as: 
\begin{equation}
\label{piparametrization}	p_1(u)=\frac{-u}{1+u+u^2}\mbox{, }\quad 	p_2(u)=\frac{1+u}{1+u+u^2}\mbox{, }\quad 	p_3(u)=\frac{u(1+u)}{1+u+u^2}\mbox{, }
\end{equation}
where, as before, the $p_i$s are arranged as $p_1<p_2<p_3$.  All possible sets of $p_i$ values are generated if $u$ is allowed to run through all values in the range $u\ge1$.

\subsection{Bianchi IX Dynamics Near the Initial Singularity}

The evolution of the Bianchi IX vacuum equations near the initial singularity was shown by Belinskii, 
Khalatnikov and Lifshitz \cite{BKL} to be well approximated by a piecewise sequence of \emph{Kasner regimes} (i.e. Kasner universes).  Within a Kasner 
regime, the $p_i$ (i.e.~$u$) values are constant and the time period over which this occurs is called an \emph{epoch}.  Successive epochs are connected by 
very brief transitions in which the value of $u$ changes from one constant to another.

When studying the Kasner regimes near the initial singularity it is convenient to reverse the time evolution so that one approaches the singularity from
a set of initial conditions associated with a time when the universe is a finite size.  Since a Kasner universe has two expanding scale factors,
and a contracting one as it evolves into the future, when travelling backwards in time toward the initial singularity, there will be {one expanding} and {two contracting} scale factors during each of the Kasner regimes. What distinguishes one epoch from another is that the expanding scale factor becomes a
contracting one and one of the contracting scale factors expands.

One can also distinguish between the two contracting scale factors.  The scale factor that decreases less rapidly is the one that exchanges roles with the increasing
scale factor upon an epoch change.  At each epoch change the scale factor that was decreasing most rapidly will continue to decrease but at a less rapid rate. Eventually
after enough epoch changes, that scale factor will decrease less rapidly than the other one and that signals an \emph{era} transition. 
On the next epoch change, the two 
decreasing scale factors exchange roles and the new era will begin.  Thus an era typically consists of a number of epochs.   

Each epoch can be associated with a particular value of $u$ and
Belinskii, Khalatnikov and Lifshitz found a simple algorithm involving the parameter, $u$, for calculating how the $p_i$ values change from one Kasner regime to the next as one ``evolves'' backwards toward the singularity. If $u>2$ the epoch changes are associated with a decrease in $u$ by precisely unity.  When
$1\leq u \leq 2$ the next subtraction of one yields the decimal part of the $u$ and inverting this signals the beginning of an era.
Thus the BKL transition rule can be put in the form of two successive iterative maps: 
\begin{eqnarray}
 & u_{N+1}=u_N-1\mbox{, } & \mbox{ if } \quad u>2 \quad \qquad \; \mbox{ (epoch change)}\nonumber\\
 & u_{N+1}=\textstyle\frac{1}{u_N-1} \mbox{, } & \mbox{ if } \quad 1\leq u\leq 2\qquad\mbox{ (era change)}
\label{uchangealgorithm} 
\end{eqnarray}
When simple decrements are performed, the epochs are in the same era.  
Clearly the smaller the decimal remainder $u_N - 1$ at the era change, the larger the number of epoch transitions in
the next era.   

\section{The Gauss Map}

Nonlinear iterative maps, such as that defined by (\ref{uchangealgorithm}), can potentially exhibit chaotic behaviour, but they can also have periodic solutions.  The ordinary epoch change, as described by the first equation in (\ref{uchangealgorithm}) does not lead to chaotic behaviour.  It is just a continual decrement of the parameter $u$ 
until its falls between one and two.  However the second part of (\ref{uchangealgorithm}) (corresponding to a simultaneous epoch and era change) {does} have a link to 
chaotic behaviour.

The equations in (\ref{uchangealgorithm}) are related to another iterative function known as the \emph{Gauss map} (see for example Rugh \cite{Rug} or Berger \cite{Ber} for more details):
\begin{equation}
u_{N+1}=\textstyle\frac{1}{u_N-[u_N]} \mbox{, }
\label{Gaussmap} 
\end{equation}
where $[u_N]$ refers to the integer part of $u_N$.  It can be assumed, without loss of generality, that all $u_i$ are positive. The Gauss map provides information only on the era changes since removing the integer part of $u$ ignores the epoch changes that occur during the successive decrements of
$u$ by unity. 

Although the Gauss map is known to be chaotic, consider the special case for which the Gauss map repeats itself. 
\begin{equation}
u_{N+1}=u_N=M+d \mbox{, }
\label{Gauss2cyclecondition} 
\end{equation}
where $M$ and $d$ (with $M\in\mathbb Z _+$ and $0\le d \le 1$), simply name the integer and decimal parts ($M=[u_N]$ and $d=u_N-[u_N]$).  Then (\ref{Gaussmap}) becomes

\begin{equation}M+d=\frac{1}{d} \nonumber \end{equation}
which leads to a simple quadratic equation for $d$ (i.e.~$ d^2+Md-1=0$) with solutions given by
\begin{equation}  d=\frac{-M\pm \sqrt{M^2+4}}{2}\mbox{. } \nonumber \end{equation}

It is clear that $M$ represents the number of epochs in each era. Choosing $M=1$ is the simplest case where each era consists of a single
epoch.  When $M=1$ the solutions for the
decimal part become
\begin{equation}
d_1=-\frac{1+\sqrt{5}}{2} \quad\mbox{ and }\quad d_2=-\frac{1-\sqrt{5}}{2}\mbox{. }
\label{GaussSolns} 
\end{equation}
Since $0\le d \le 1$, only the second value is a valid solution.  Therefore, a solution to the periodic Gauss map is
\begin{equation}
u_N=M+d_2=1+(-\frac{1-\sqrt{5}}{2})=\frac{1+\sqrt{5}}{2}\mbox{,}
\end{equation}
which is the golden ratio $\phi$. The solution that was rejected is just $-\phi$.
For $M>1$ the Gauss map is also repetitive and the solutions to the resulting quadratic equation are the so-called ``silver ratios''. 

The link between the Gauss map and the golden ratio is well known, as is the link between the era transitions and the Gauss map.  This provides a connection 
between transitions in the BKL approximations to the Bianchi IX dynamics and the golden ratio. However we will show that the connection is much stronger.  
As will be seen 
in subsequent sections, the use of Ellis-MacCallum-Wainwright variables \cite{EM,WH}, leads to a method where the golden ratio can be derived directly from
the full Einstein equations without having to rely on the reduction to an iterative map.

\section{The Golden Ratio}\label{GRatio}

Interest in the golden ratio has had a long history and it has many intriguing properties. Recently published books on the subject have ranged from the popular \cite{Liv} to
the more mathematically inclined \cite{PL}.  
The purpose of this section is to provide a brief review of some properties of the golden ratio that will be relevant to the discussion to follow.  Given a line segment, $\overline{ACB}$, $C$ divides $\overline{AB}$ into a golden ratio if, in terms of measures, $\frac{\overline{AC}}{\overline{CB}}=\frac{\overline{AB}}{\overline{AC}}$.  In other words, the ratio of the larger segment to the smaller segment, equals the ratio of the whole to the larger segment.  Since only ratios are considered, the overall scale is not important and can be set so that $\overline{AC}=x$, and $\overline{CB}=1$.

The proportion then becomes,
\begin{equation} \frac{x}{1} =\frac{(x+1)}{x} \end{equation}
which leads to the quadratic equation:
\begin{equation} \label{goldquadratic} x^2-x-1 =0 \end{equation}
The solutions are,
\begin{equation}
\label{goldquadsolns} x_1=\frac{1+\sqrt{5}}{2}\quad \mbox{ and }\quad x_2=\frac{1-\sqrt{5}}{2}\mbox{.}
\end{equation}
The first solution is the golden ratio, $\phi = \frac{1+\sqrt{5}}{2}=1.6180339887\cdots$.  It has two unusual (self-referential) properties.

Firstly, the value of its inverse is equal to one less than its value,
\begin{equation}
\label{phiinverse}\frac{1}{\phi}=\phi-1
\end{equation}
and secondly, the value of its square is one more than its value, 
\begin{equation}
\label{phisquared}\phi^2=\phi+1\mbox{,\quad	(using (\ref{goldquadratic})) }
\end{equation}
Note that from (\ref{phiinverse}), 
\begin{equation}
\label{negphiinv} \textstyle -\frac{1}{\phi}=1-\phi=\frac{1-\sqrt{5}}{2}\mbox{,}
\end{equation}
which in (\ref{goldquadsolns}) is the second solution of (\ref{goldquadratic}).  Finally a \emph{golden rectangle} is defined as a rectangle that has its length-over-width ratio equal to $\phi$.  

\section{The EMW Approach}

Returning to the dynamics of the Bianchi IX cosmologies, we will take an alternative approach. This method was 
introduced by Ellis and MacCallum who used an orthonormal tetrad technique to give the dynamical equations in terms of the spatial curvature, the shear and the
Hubble expansion function.  Wainwright subsequently normalized the curvature and shear with respect to the Hubble expansion which provided a better understanding of 
the dynamics of the anisotropies. We shall call this approach that of    
Ellis-MacCallum-Wainwright or EMW \cite{EM, WH}.

The phase-space variables in the EMW formalism are $(\Sigma_+,\Sigma_-,N_1,N_2,N_3 )$, where $\Sigma_+$ and $\Sigma_-$ are the components of a traceless shear tensor, 
and the $N_i$ are spatial curvature components (all are dimensionless and expansion-normalized).  A normalized time parameter, $\tau$, is also introduced so that the Einstein equations take the following form:
\begin{eqnarray}
N_1' & = & (q-4\Sigma_+ ) N_1 \nonumber \\
N_2' & = & (q+2\Sigma_++2\sqrt{3} \Sigma_- ) N_2 \nonumber \\
\label{Nieqns}N_3' & = & (q+2\Sigma_+-2\sqrt{3} \Sigma_- ) N_3 \\
 \Sigma_+' & = & -(2-q) \Sigma_++3S_+ \nonumber \\
 \Sigma_-' & = & -(2-q) \Sigma_-+3S_- \nonumber
\end{eqnarray}

Here the prime ($\prime$) indicates a derivative with respect to $\tau$, and
\begin{eqnarray}
S_+ & = & \textstyle \frac{1}{2} [(N_2-N_3 )^2-N_1 (2N_1-N_2-N_3 )]\mbox{, } \nonumber \\
S_- & = & \textstyle \frac{\sqrt{3}}{2} [(N_3-N_2 )(N_1-N_2-N_3 )]\mbox{, } \nonumber \\
\label{otherparams}q & = & \textstyle \frac{1}{2} (3\gamma-2)(1-K)+\frac{3}{2}(2-\gamma)\Sigma^2\mbox{, } \\
\Sigma^2 & = & \textstyle \Sigma_+^2+\Sigma_-^2\mbox{, } \nonumber \\
K & = & \textstyle \frac{3}{4} [N_1^2+N_2^2+N_3^2-2(N_1 N_2+N_2 N_3+N_3 N_1 )]\mbox{, } \nonumber
\end{eqnarray}
and the constant, $\gamma$, is obtained from the perfect fluid equation of state:
\begin{equation}
\label{perfectfluid}p=(\gamma-1)\rho
\end{equation}
where $p$ and $\rho$ are, respectively the pressure and mass density of the matter in the spacetime.  

When constructing numerical solutions to the system of ODEs (\ref{Nieqns}), the curvature variables $N_i$ ($i=1,2,3$) present challenges as they approach the initial 
singularity. Two components always tend toward zero, while the other one grows.  Numerical accuracy
can be lost very quickly and the computer program either crashes or leads to incorrect solutions. 
This problem can be avoided by transforming from the $N_i$s to a new set of variables $Z_i$s defined by,
\begin{equation}
\label{Zidefn} N_i=\exp(-Z_i )
\end{equation}
and making the change $\tau\rightarrow-\tau$.  The new equations are
\begin{eqnarray}
Z_1' & = & q-4\Sigma_+ \nonumber \\
Z_2' & = & q+2\Sigma_++2\sqrt{3} \Sigma_- \nonumber \\
\label{Zieqns}Z_3' & = & q+2\Sigma_+-2\sqrt{3} \Sigma_- \\
\Sigma_+' & = & (2-q) \Sigma_++3S_+ \nonumber \\
\Sigma_-' & = & (2-q) \Sigma_-+3S_- \nonumber
\end{eqnarray}

It is useful to understand the system dynamics from the perspective of the new variables.  Recall that the piecewise Kasner approximation to the Bianchi IX universe is valid near the singularity.  Its metric is given in (\ref{KasnerMetric}), thus (\ref{eqnsforpis}) holds in each Kasner regime.

Expressing the $p_i$s in terms of the variables in (\ref{Zieqns}) produces \cite{WH}
\begin{eqnarray}
p_1 & = & \frac{1}{3}(1-2 \Sigma_+ ) \nonumber \\
\label{piinSigma} p_2 & = & \frac{1}{3}(1+ \Sigma_++\sqrt{3}\Sigma_- ) \\
p_3 & = & \frac{1}{3}(1+ \Sigma_+-\sqrt{3}\Sigma_- ) \nonumber
\end{eqnarray}
and inserting these into (\ref{eqnsforpis}) yields
\begin{equation}
\label{Kasnercondition} \Sigma^2= \Sigma_+^2+ \Sigma_-^2=1
\end{equation}
which will be referred to as the \emph{Kasner condition}.  Equation (\ref{Kasnercondition}) defines the unit circle centered on the origin of the  $\Sigma_+\Sigma_-$-plane, and this circle will be referred to as the \emph{Kasner ring}.  Since the Kasner solution is spatially flat with $N_1=N_2=N_3=0$, all possible Kasner solutions always lie on the Kasner ring in the five-dimensional phase space.

For arbitrary initial conditions the Kasner condition in general will not hold nor will it hold during the brief transitions between epochs, however as the spacetime 
approaches the singularity it will hold asymptotically since $\Sigma^2\rightarrow 1$ and $N_i\rightarrow 0$ as $\tau \rightarrow \infty$ (i.e. as $t\rightarrow0$).

Applying the above conditions to (\ref{otherparams}), leads to $K\rightarrow0$, and $S_\pm\rightarrow0$ as $\tau\rightarrow\infty$.  Thus in this limit, the parameter $q$ becomes
\begin{eqnarray}
& q & = \textstyle\frac{1}{2}(3\gamma-2)(1-K)+\frac{3}{2}(2-\gamma)\Sigma^2 \nonumber \\ 
&	& = \textstyle\frac{3}{2}\gamma-1+3-\frac{3}{2} \gamma \nonumber \\
\label{qeq2}&   & = 2\mbox{ .}
\end{eqnarray}

This means that as the spacetime approaches the initial singularity, $q$ is independent of $\gamma$.  Since $\gamma$, (through $q$), provided the only 
matter-dependence in Einstein's equations ((\ref{Nieqns}) or (\ref{Zieqns})), this justifies the assertion, made earlier, that the properties of matter are unimportant to the dynamics of this universe close to the initial singularity. 
\begin{figure}
\begin{center}
\vspace{-0.4cm}
\includegraphics[width=7.7cm]{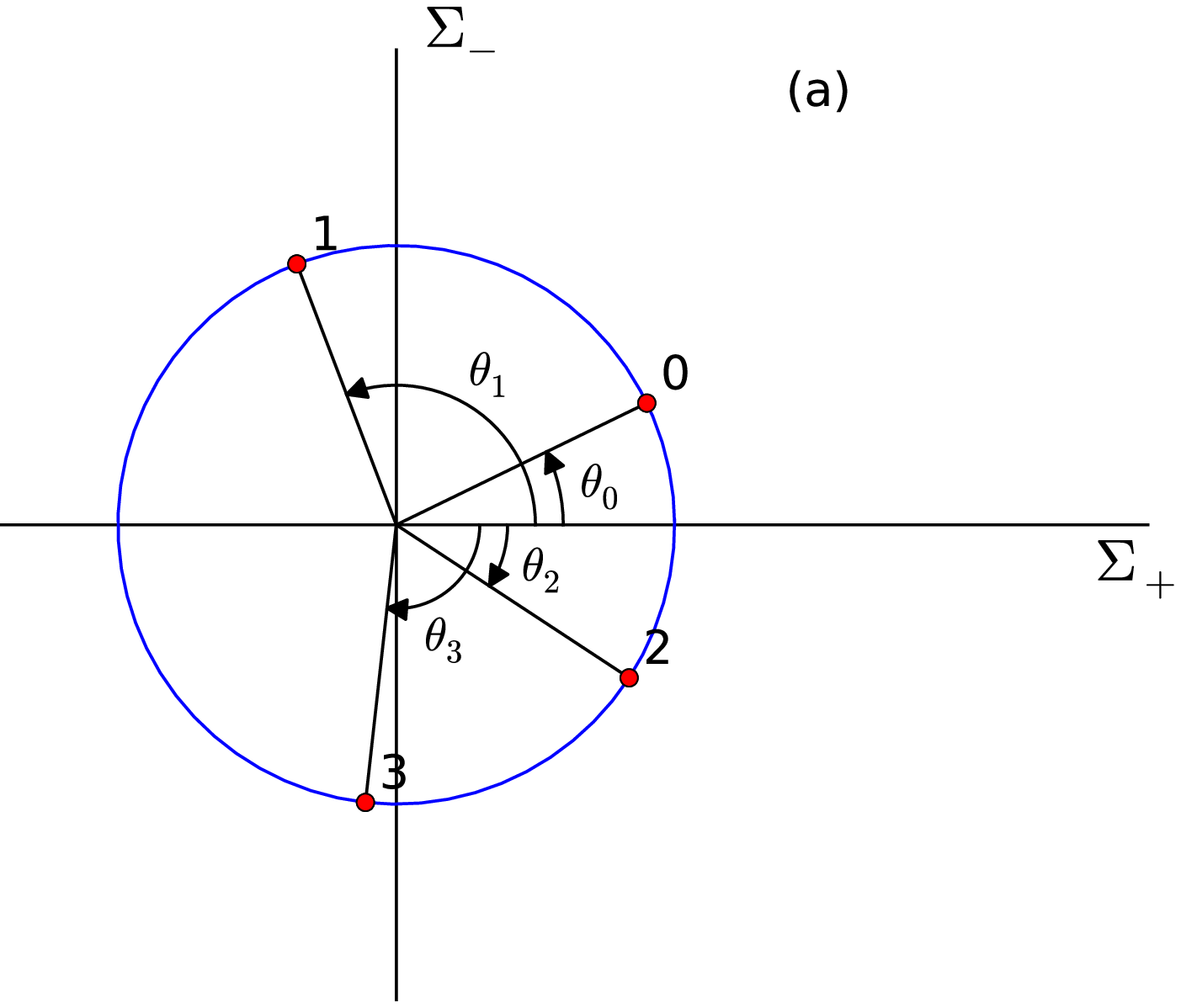}
\includegraphics[width=7.7cm]{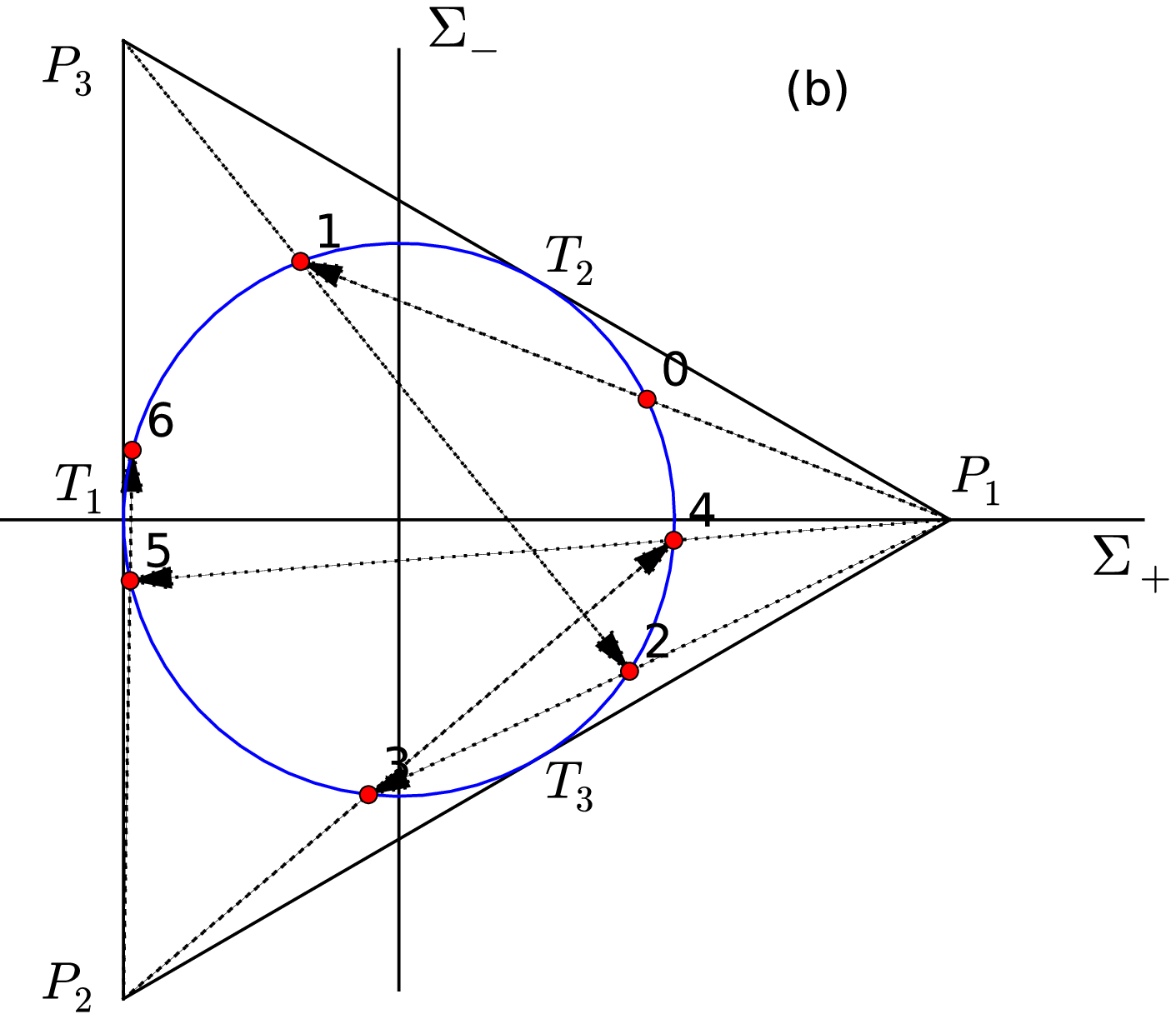}
\caption{(a) Action of the B-map given by (\ref{Bmap}). (b)  A geometric construction of the B-map.} \label{BmapIntro}
\vspace{-0.6cm}
\end{center}
\end{figure}
\section{The B-Map}
Recall that in the BKL formulation, on the approach toward the initial singularity, the evolution of the dynamics for the Bianchi IX cosmology can be considered as a piecewise function, composed of linear regimes called epochs that were grouped into eras.  

$\Sigma_+$ and $\SigmaΣ_-$ are related to the $p_i$s from the BKL approach (as seen in (\ref{piinSigma})). An epoch (in the EMW approach) corresponds to the
cosmology evolving close to a vacuum Kasner system with $N_i \approx 0$ and $\Sigma^2 \approx 1$.  In the five-dimensional phase space of the EMW dynamical variables, the
system corresponds to one where the solution is located on the Kasner ring until a transition occurs. The transition from one point on the Kasner ring to
another position on the ring, occurs when the parameter $u$ changes, leading to changes in the $p_is$ which in turn lead to changes in $\Sigma_\pm$. 
\begin{figure}[h]
\begin{center}
\vspace{-0.4cm}
\includegraphics[width=8.0cm]{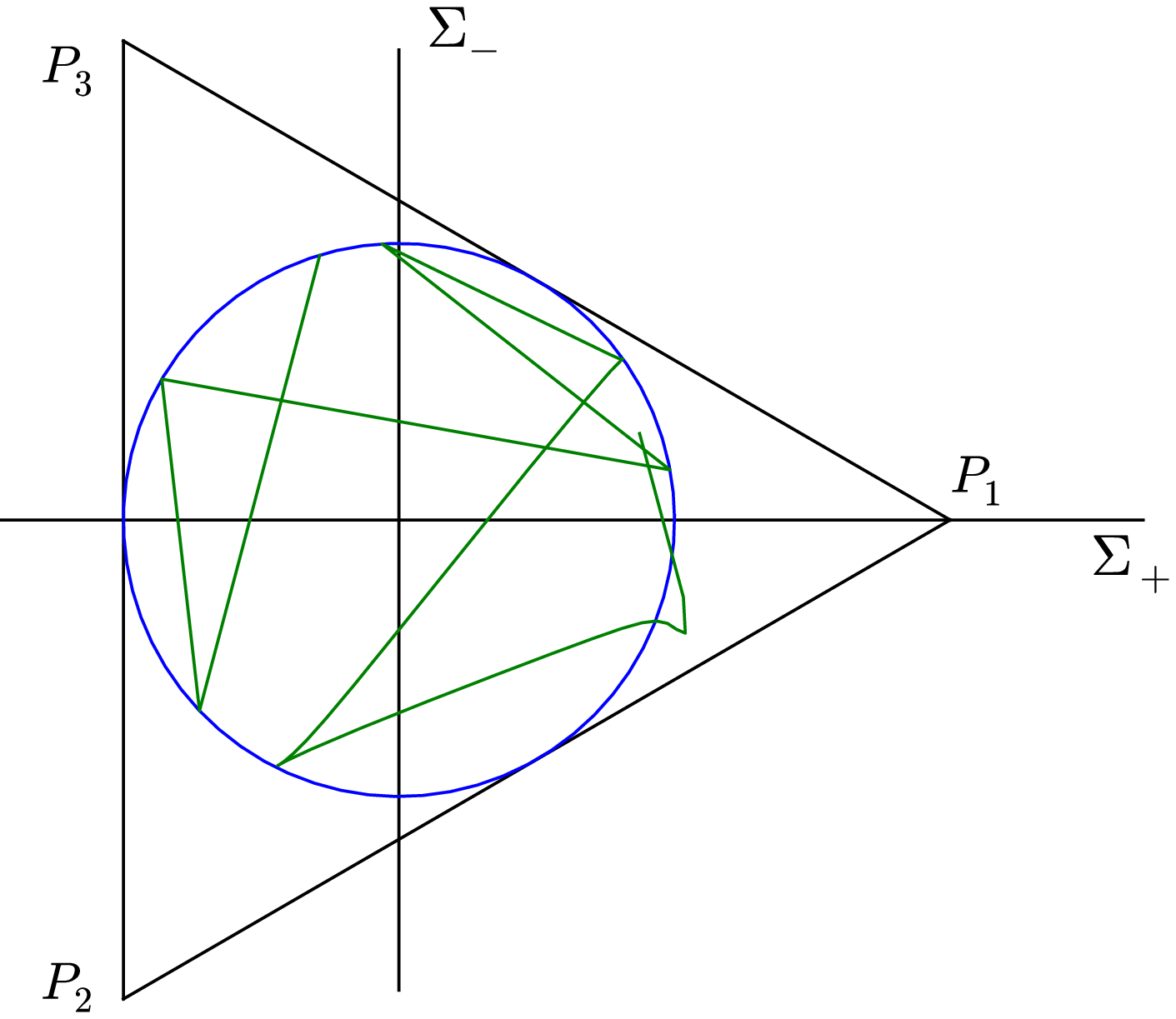}
\caption{A ($\Sigma_+$,$\Sigma_-$) plot of a simulation run solving Einstein's equations, showing the transient phase and then B-map convergence. The initial
conditions for this simulation are $Z_1 = -\ln(0.2)$, $Z_2 = -\ln(3.1)$, $Z_3 = -\ln(2.9)$, $\Sigma_+ = .90$, $\Sigma_- = .25$} \label{Convergence_to_Bmap}
\vspace{-0.0cm}
\end{center}
\end{figure}

\begin{figure}[h]
\begin{center}
\vspace{-0.1cm}
\includegraphics[width=7.7cm]{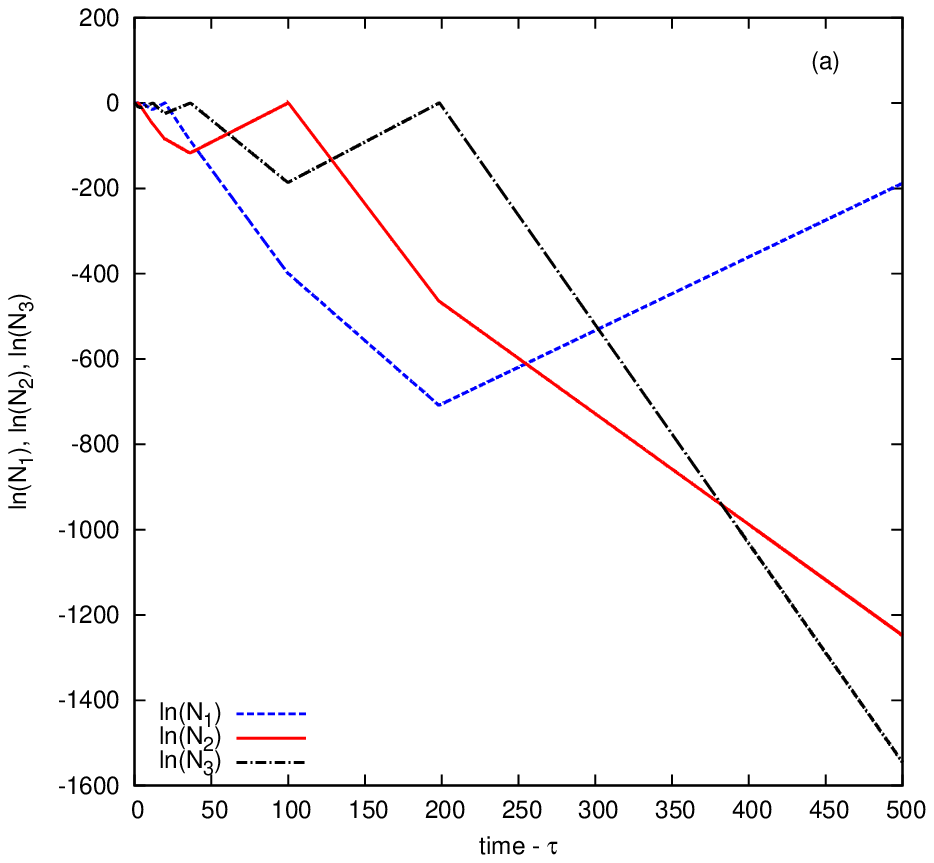}
\includegraphics[width=7.7cm]{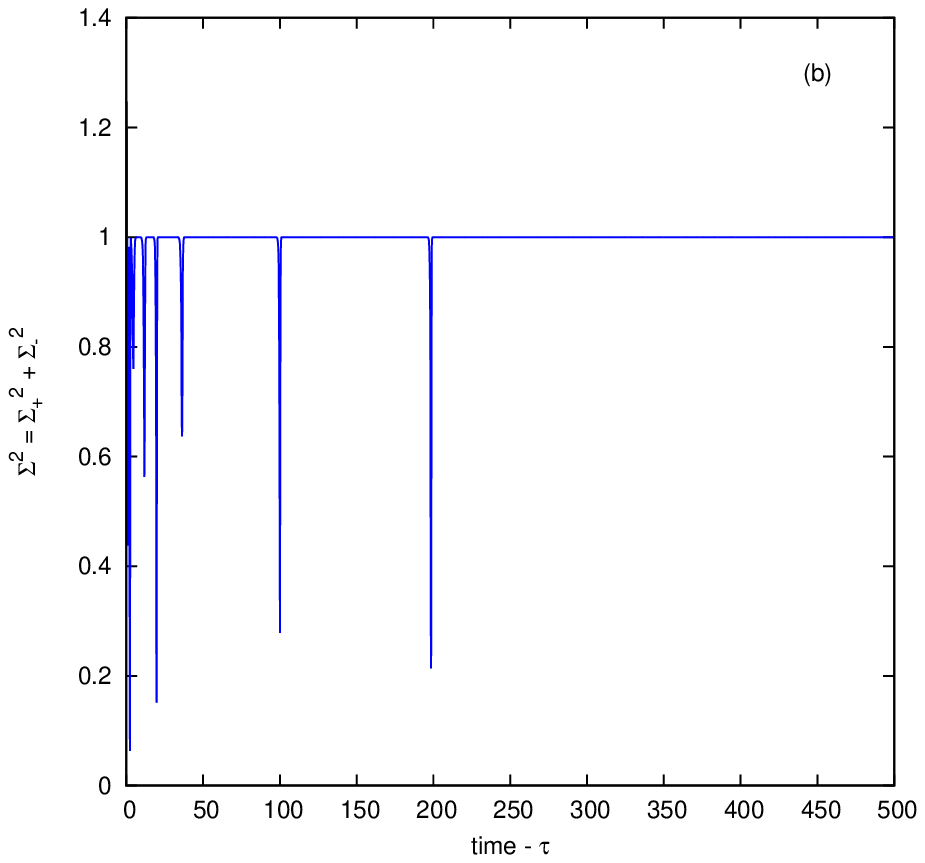}
\vspace{-0.2cm}
\caption{A time series representation of the same simulation as in Figure \ref{Convergence_to_Bmap}.  (a) The $\ln(N_i)$ vs $\tau$ graph.  (b) The $\Sigma^2=\Sigma_+^2+ \Sigma_-^2$ vs $\tau$ graph.} \label{ZiandSigmaConvergent}
\vspace{-0.8cm}
\end{center}
\end{figure}

Bogoyavlensky \cite{Bog} interpreted these transitions as an iterative map from the Kasner ring onto itself, and found the equation for the map.  This will be referred to as the \emph{B-map} and can be described as follows:
Given an epoch, represented by an angle $\theta_n$ (measured in radians from the appropriate axis), the subsequent epoch,  $\theta_{n+1}$, can be found by using,
\begin{equation}
\label{Bmap}\theta_{n+1}=\arccos(\frac{4-5\cos\theta_n}{5-4 \cos\theta_n}) \mbox{, }\quad  0\le\theta_n\le \frac{\pi}{3}
\end{equation}
when $\theta_n$ is in the given range.  For angles in other ranges, the formula is reflected in the $\Sigma_+$-axis, and/or rotated by $\frac{2\pi}{3}$ once or twice.  
Figure \ref{BmapIntro}(a) provides an example of the action of the iterative map, (\ref{Bmap}) on the transitions from one angular position to another on
the Kasner ring. 
Wainwright and Hsu \cite{WH} clarified the B-map by representing it as a geometric construction.  Consider the Kasner ring on the $\Sigma_+ \Sigma_-$-plane and draw an equilateral triangle with sides tangent to the Kasner ring and with one vertex on the $\Sigma_+$-axis, as shown in Figure \ref{BmapIntro}(b).

Consider the point labelled 0 in Figure \ref{BmapIntro}(b) as the initial point of the
B-map dynamics.  Except for the points that are tangent to the equilateral triangle, each point on the Kasner ring will have a triangle vertex closer to it compared to the
other two vertices. Take the equilateral triangle vertex which is closest to the point (in this case $P_1$) and construct a line that passes through the closest vertex and the point on the Kasner ring.  That line can be extended to intersect another point on the Kasner ring (in this case point 1). This point represents the new
Kasner solution that follows after the Bianchi IX transition. The procedure then continues indefinitely or until the solution hits one of the tangent points $T_i$.
Each of these points is mapped onto itself since they are equidistant from the two nearest vertices of the equilateral triangle $\Delta P_1P_2P_3$. Thus these points are equilibrium
points of the B-map.

In a Bianchi IX cosmology with arbitrary initial conditions, the B-map is an approximation to the actual dynamics of the system; one that gets better and better as the singularity is approached. Figure \ref{Convergence_to_Bmap} shows a plot of the evolution of $\Sigma_+$ and $\Sigma_-$ during a simulation that employed a fourth-order,
 variable-step Runge-Kutta method to solve the differential equations (\ref{Zieqns}).  It can be seen that initially, in the transient phase, the values do not match any B-map dynamics, 
but that it quickly converges to one as $\tau$ increases. Figure \ref{ZiandSigmaConvergent} shows a time series for the $\ln(N_i)$, and $\Sigma^2=\Sigma_+^2+ \Sigma_-^2$ 
obtained
from the same simulation. The plots indicate that the normalized spatial curvatures $N_i$ tend toward zero while $\Sigma^2$
is very well approximated by unity during each epoch.  These plots, as well as the transitions on the $\Sigma_+\Sigma_-$-plane, represent the generic behaviour of 
Bianchi IX cosmologies close to the singularity. 
One easily sees that each era can consist of a number of epochs and that the changes in epochs are associated with a significant dip in the value of
$\Sigma^2$ which will be quantified later.  

\subsection{The B-Map: Epochs and Eras}

\begin{figure}
\begin{center}
\vspace{-0.4cm}
\includegraphics[width=7.0cm]{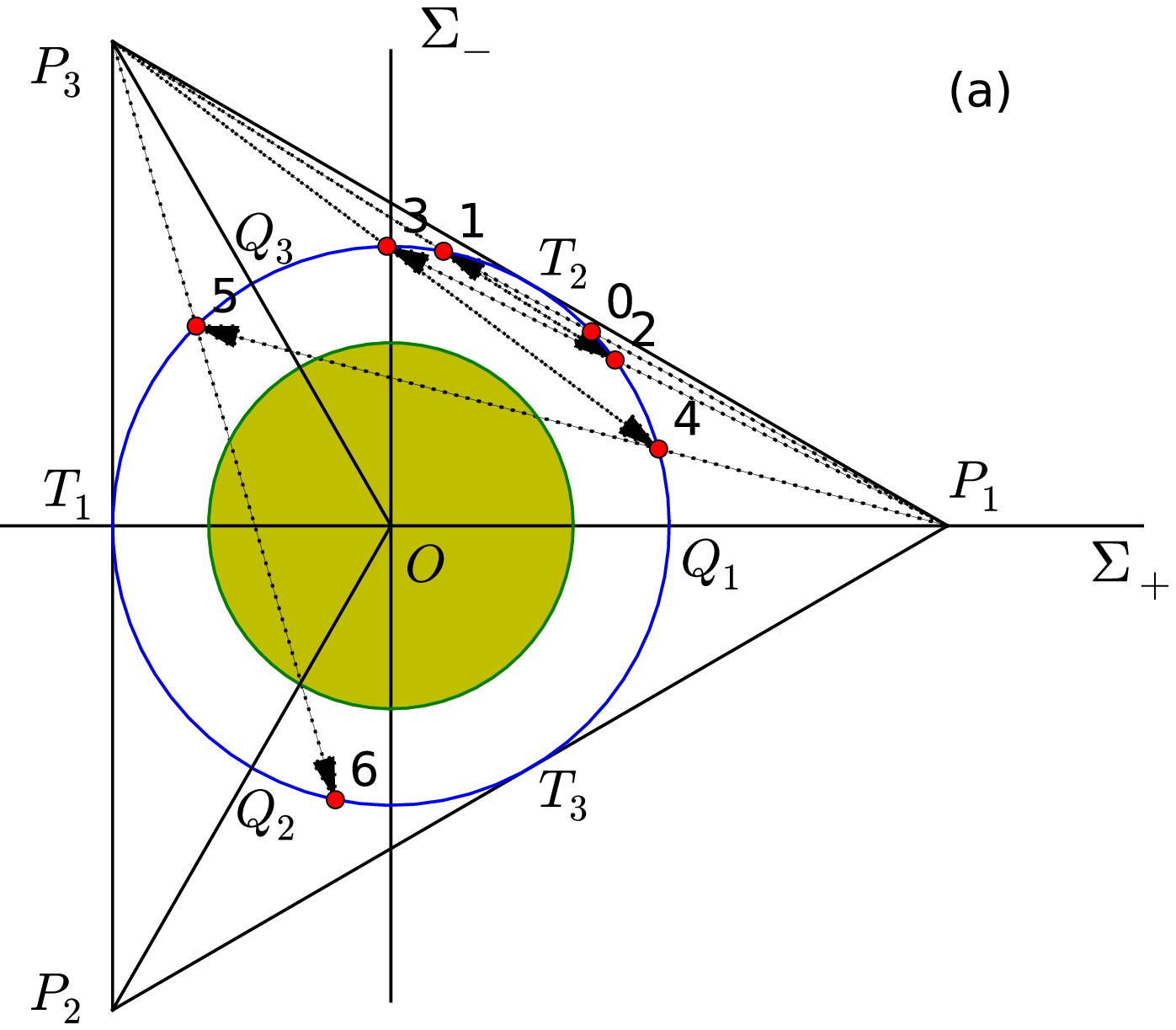}
\includegraphics[width=6.2cm]{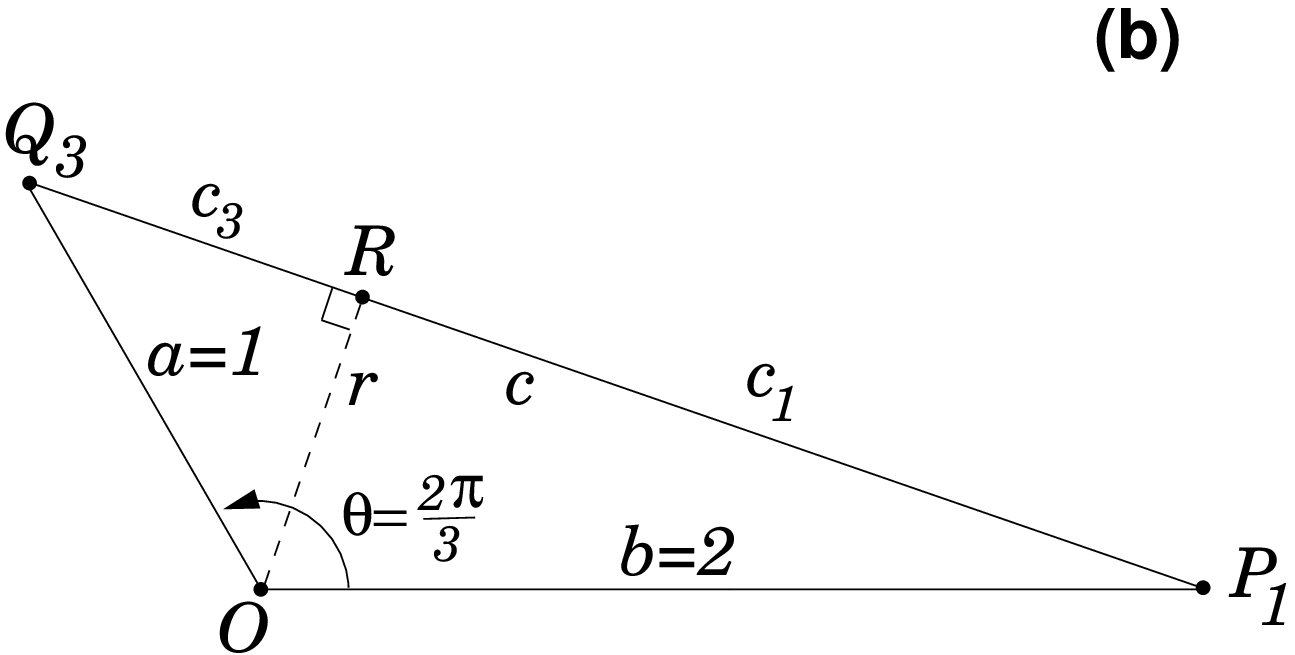}
\caption{(a) Within the B-map, an era change occurs whenever a transition line passes within the shaded circle of radius $r =\sqrt{{3}/{7}} \approx 0.654654$. (b) The triangle shows how that limiting radius is is computed geometrically.} \label{BmapEra}
\vspace{-0.6cm}
\end{center}
\end{figure}

While the full dynamical behaviour involving the curvature variables is is not captured by the B-map, one can use the geometry of the map to determine the conditions 
required for a transition between eras.
Returning to the Kasner ring and its circumscribed equilateral triangle (See Figure \ref{BmapEra}(a)), epoch changes 
that that do not cause an era change are those that occur between two
curvature variables which exchange roles as increasing and decreasing modes. This means that the B-map oscillates around one of the equilibrium points $T_i$.

Figure \ref{BmapEra}(a) also shows that the Kasner ring can be broken into three equal arcs $\wideparen{Q_1T_2Q_3}$, $\wideparen{Q_3T_1Q_2}$, and 
$\wideparen{Q_2T_3Q_1}$ - i.e.~those lying inside $\Delta P_1OP_3$, $\Delta P_3OP_2$ and $\Delta P_2OP_1$, respectively. Thus the
lines connecting the origin with the three vertices of the triangle act as boundaries to those three regions.  In Figure \ref{BmapEra}(a) the first four transitions
all occur inside $\Delta P_1OP_3$, the fifth transition crosses the boundary $\overline{OP_3}$ which ends the oscillation between the vertices $P_1$ and $P_3$.  A transition 
from one triangle to another signals the change of an era.  In the figure, the limiting case occurs when a point on the Kasner ring is mapped 
onto the intersection of the line segment $\overline{OP_i}$ with the Kasner ring at a point labelled $Q_i$.  That point gets mapped in the next iteration onto an equilibrium point $T_i$. In the example shown in Figure \ref{BmapEra}(b) the geometric construction
of the limiting case creates a $\Delta P_1OQ_3$ where $Q_3$ is the point of intersection between the Kasner ring and the line segment $\overline{OP_3}$.  

The line segment $\overline{P_1Q_3}$ has a distance of closest approach to the point $O$ (here called $r$ or the length of $\overline{OR}$). A circle with
a radius $r$ then acts as a boundary
that defines when an era change occurs during a Kasner epoch transition. The radius $r$ 
can be found using the triangles in Figure \ref{BmapEra}(b) by the following argument:
Given the construction of the Kasner ring and the equilateral triangle $\Delta P_1P_2P_3$, $a=1$ is the length of $\overline{OQ_3}$ and $b=2$ is the length of $\overline{OP_1}$.  Also from the construction of $\overline{OP_1}$, $\angle Q_3OP_1 = \frac{2\pi}{3}$ so that by the cosine rule for triangles the third side of $\Delta Q_3OP_1$ has length $c = \sqrt{7}$.  Sub-divide $\Delta Q_3OP_1$ into $\Delta Q_3OR$ and $\Delta P_1OR$ such
that $\overline{OR}$ $\perp$ $\overline{P_1Q_3}$.  Let the length of $\overline{Q_3R}$ be $c_3$, the length of $\overline{P_1R}$ be $c_1$. 
This leads to three equations for the three unknowns $c_1$, $c_3$ and $r$:
\begin{equation}  c_1 + c_3 = \sqrt{7}; \qquad \qquad 1 = r^2 + c_3^2; \qquad \qquad 4 = r^2 + c_1^2 . \nonumber \end{equation}
Solving for $r$ one obtains
\begin{equation} r = \sqrt{\frac{3}{7}}. \nonumber \end{equation}
Therefore any transition between Kasner epochs described by a B-map line segment passing inside a circle of
radius $r = \sqrt{3/7}$ is also associated with an era transition.  
Figure \ref{BmapEra}(a) shows a shaded circle of radius $ r= \sqrt{{3}/{7}}$. The epochs represented by points 0 to 4 occur within the first era, the second era has only a single epoch, corresponding to point 5, and the third era begins with the epoch corresponding to point 6. In addition, a time series plot for $\Sigma^2$ can be used to
indicate an era change when $\Sigma^2 \leq 3/7$.  

\subsection{Kasner transition dynamics}
The B-map does not provide all of the information regarding the Bianchi IX dynamics since it only deals with transitions in the
$\Sigma_+ \Sigma_-$-plane where all the $N_i$s vanish.  However it has been shown that the transitions between Kasner solutions are governed by Bianchi II Taub vacuum
solutions \cite{Taub}. The two-parameter Taub solution written in so-called ``Kasner form'' (See the discussion in \cite{WE}) has a line element given by:
\begin{equation}
\label{TaubMetric}ds^2=A^2dt^2-t^{2p_1}A^{-2}(dx + 4p_1 b z dy)^2-t^{2p_2}A^2dy^2-t^{2p_3}A^2dz^2
\end{equation}
with
\begin{equation}
\label{eqnsfortaub}A^2 = 1 + b^2 t^{4 p_1}, \quad  \quad p_1+p_2+p_3=1\quad \quad p_1^2+p_2^2+p_3^2=1\mbox{. }
\end{equation}
When the parameter $b=0$ the solution reduces to the Kasner case.
When $b\neq 0$ one and only one of the $N_is$ (depending on the choice of $u$) is non-zero.  Therefore the dynamics of a transition takes place in a 3D subspace of the
full 5D phase space.  The Taub transitions follow paths on an ellipsoidal surface defined by :\begin{equation}
\Sigma_+^2+\Sigma_-^2+\textstyle\frac{3}{4}N_i^2=1 \label{ellipsoideqn}
\end{equation}
where $N_i$ is the particular dimensionless spatial curvature variable which is non-zero during the transition. Here it should be noted that
the factor of $\frac{3}{4}$ used here is consistent with our normalization (and that of Ma and Wainwright \cite{MW}). For a different normalization
see \cite{WE}.

\begin{figure}
\begin{center}
\includegraphics[width=7.7cm]{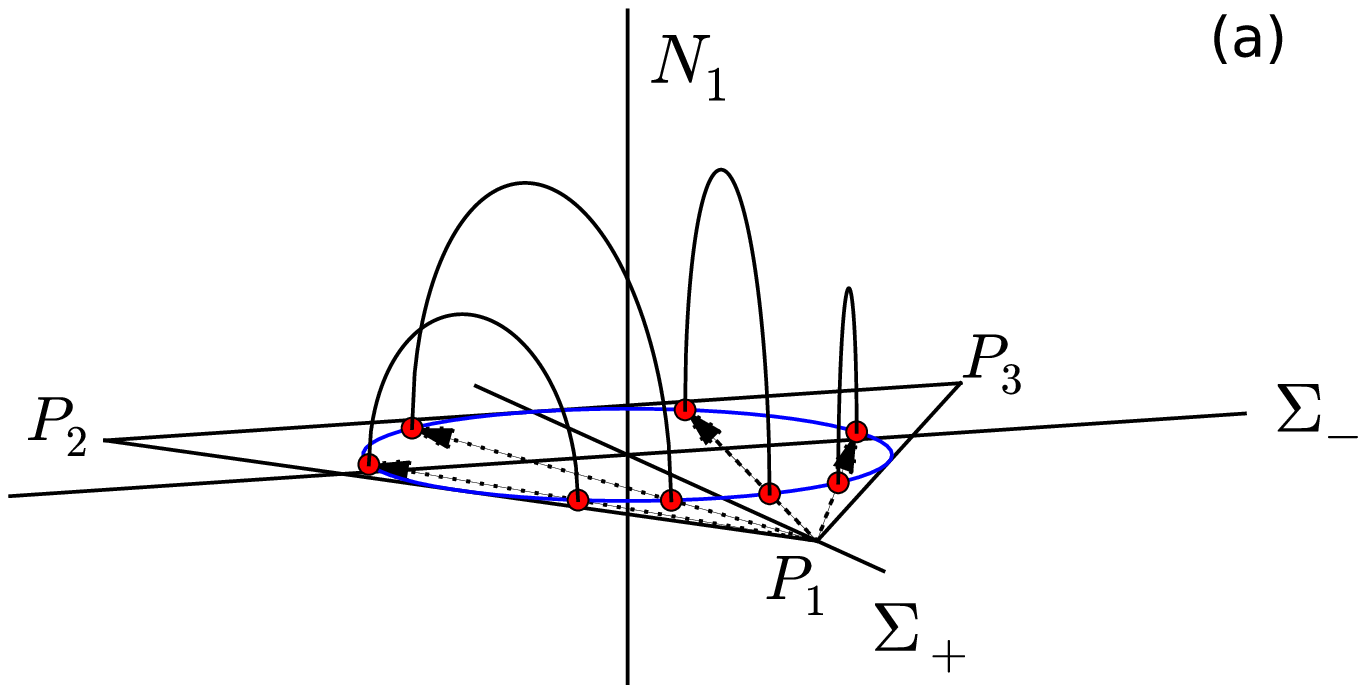}
\includegraphics[width=7.7cm]{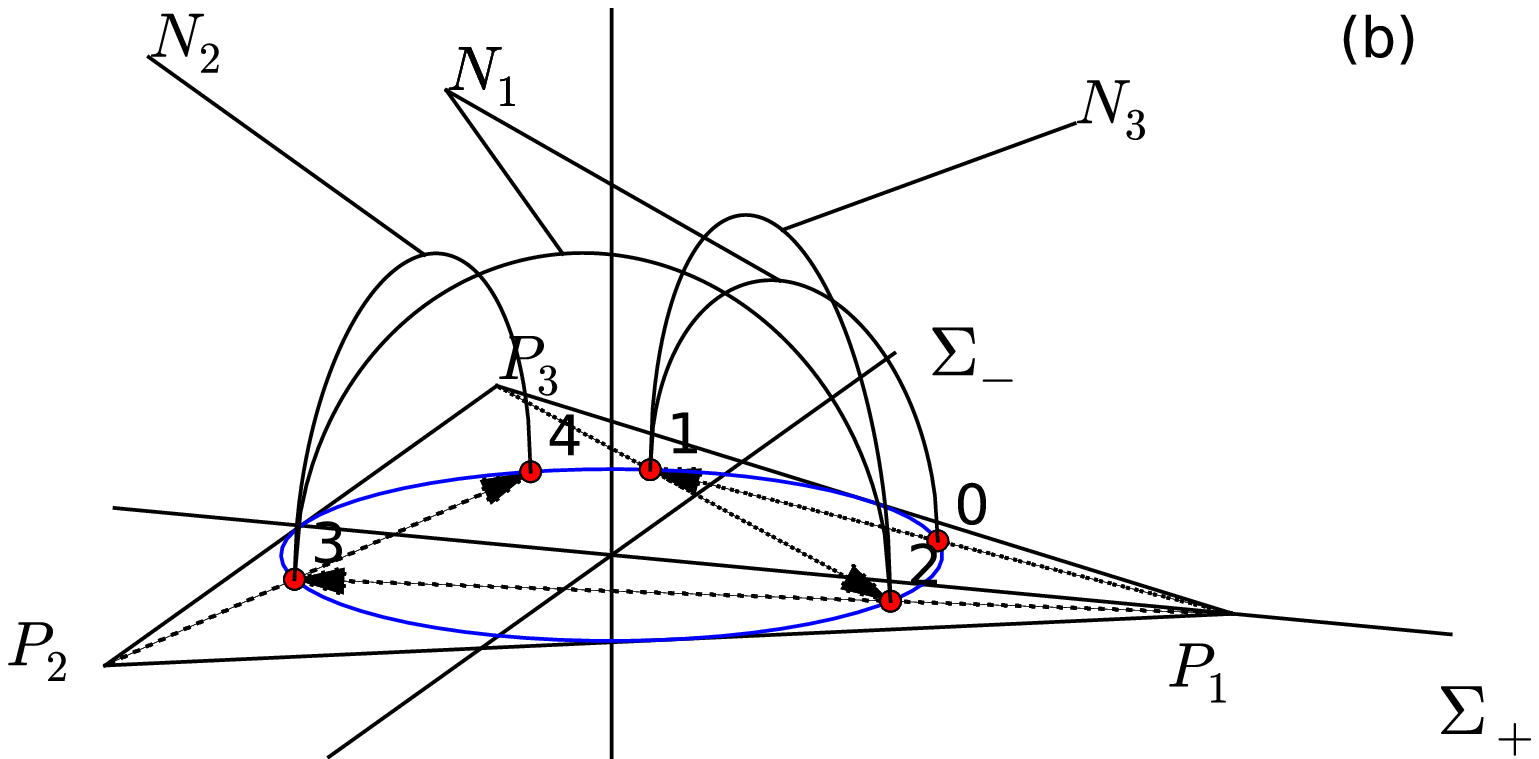}
\caption{Transitions represented by projections from $P_i$ have $N_i$ as the only non-zero curvature variable and trace out curves on the surface of an ellipsoid.  (a) Various
projections from $P_1$, allowing the ellipsoid to be visualized. (b) A B-map sequence showing the elliptical path of each non-zero $N_i$ variable.} \label{Ellipses}
\vspace{-0.6cm}\end{center}
\end{figure}

In the Figure \ref{BmapIntro} transition from  point 0 to point 1, it is the variable $N_1$ which is non-zero.  $N_1$ will be the non-zero curvature variable for any transitions
 that are
projections from vertex $P_1$.  Figure \ref{Ellipses}(a) shows some of these, revealing the ellipsoid structure. Similarly, $N_2$ and $N_3$ are the non-zero $N_i$s for B-map transitions that are projections from points $P_2$ and $P_3$, respectively (see Figure \ref{Ellipses}(b)). The transitions in the $\Sigma_+ \Sigma_-$-plane are simply the projections of the Taub orbits onto the $\Sigma_+\Sigma_-$-plane. As the $N_i$ evolves away from zero, the point ($\Sigma_+$, $\Sigma_-$) moves into the
interior of the Kasner ring and only returns to the ring when the curvature variable once again vanishes.

\section{The 3-Cycle}

\begin{figure}
\begin{center}
\vspace{-0.4cm}
\includegraphics[width=7.7cm]{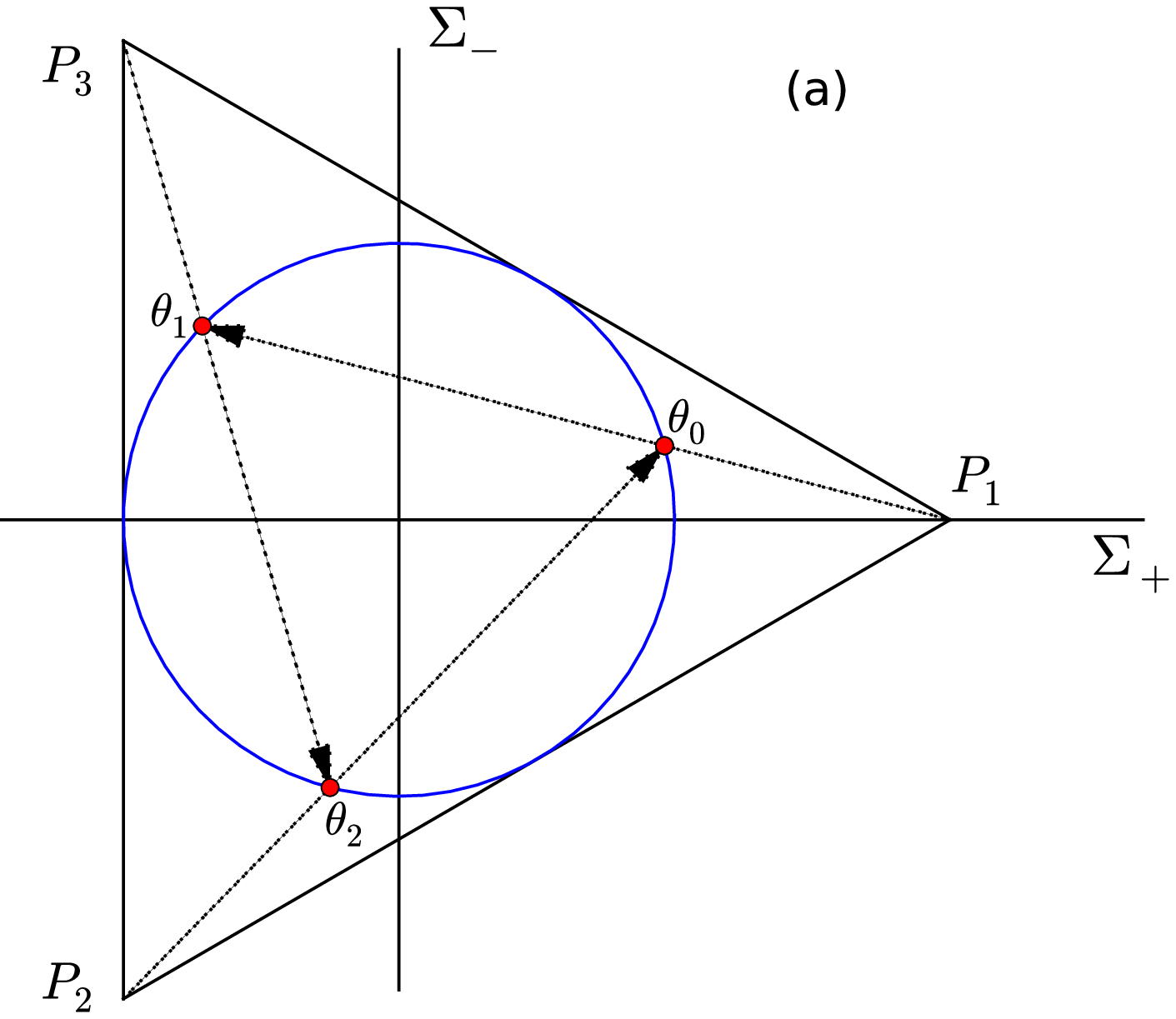}
\includegraphics[width=7.7cm]{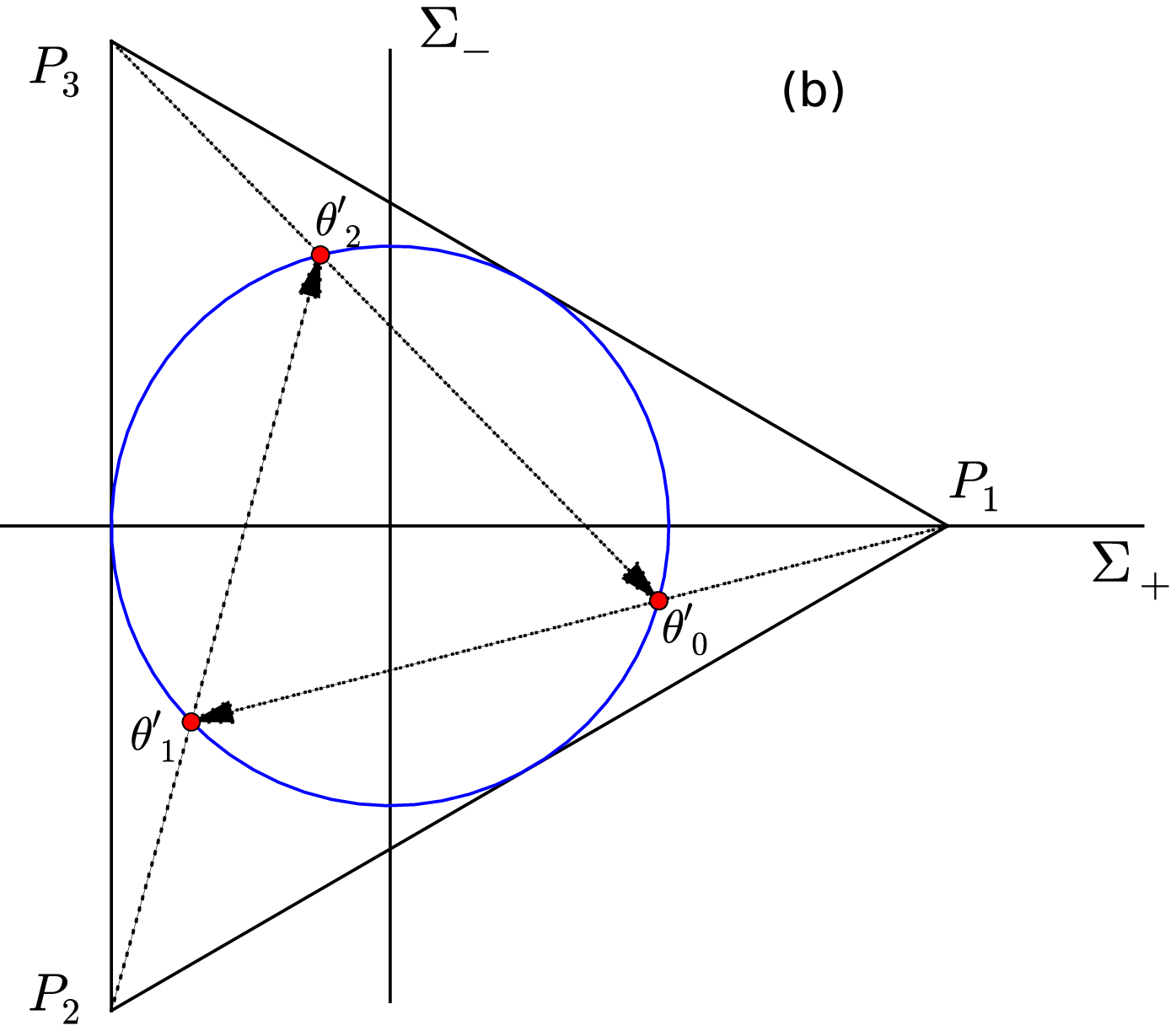}
\vspace{-0.4cm}
\caption{(a) The counterclockwise 3-cycle. (b) The clockwise 3-cycle.} \label{Bmap3cycle}
\vspace{-0.6cm}
\end{center}
\end{figure}

The B-map loses information in each transition and thus can be chaotic \cite{Bog}. That this is so is due to the fact that a point (other than one of the $T_i$s) on the Kasner circle could
have come from one of two prior Kasner states. 
While generic initial conditions lead to chaotic orbits across the Kasner ring, there are also an infinite number of periodic orbits. These orbits will proceed around
the ring and after an integer number of transitions will return to the initial position on the ring.  
The simplest such orbits are where each era consists of a single epoch and these are
shown in Figure \ref{Bmap3cycle}. There are only two possibilities in this case depending on whether the transitions follow a clockwise or counter-clockwise
pattern around the Kasner ring.  It is also of interest to note that the existence of a 3-cycle in iterative maps is sufficient to prove 
that the map is chaotic \cite{LY}. 

\subsection{The 3-Cycle: Identifying the Kasner States}\label{3CyclePoints}

The 3-fold symmetry of the periodic B-map shown in Figure \ref{Bmap3cycle} implies that the Kasner states are separated one from the other by an
angular displacement
$\Delta \theta = \frac{2\pi}{3}$ measured from the origin.  The points on the counterclockwise 3-cycle will be labelled as $\theta_0$, $\theta_1$, and $\theta_2$.  The B-map equation (\ref{Bmap}) must hold, 
and it must map the angle, $\theta_0$, to $\theta_1 = \theta_0+\frac{2\pi}{3}$.  Therefore 
\begin{equation} \cos(\theta_0+{\textstyle\frac{2\pi}{3}})  =  \frac{4-5\cos\theta_0}{5-4 \cos\theta_0}. \nonumber \end{equation}
Using the trigonometric identity for the cosine of the sum of two angles leads to:
\begin{equation} \cos \theta_0 \cos {\textstyle{\frac{2\pi}{3}}} - \sin \theta_0 \sin {\textstyle{\frac{2\pi}{3}}}
= \frac{4-5\cos\theta_0}{5-4 \cos\theta_0} . \nonumber \end{equation} 
Replacing the trigonometric functions of $2\pi/3$ with their numerical values, substituting $\sqrt{1-\cos^2\theta}$ for $\sin \theta $, squaring the result and collecting
common powers of $\cos \theta$ leads to a quartic equation for $\cos \theta_0$:
\begin{equation}  64\cos^4\theta_0-80\cos^3\theta_0-12\cos^2\theta_0+40\cos\theta_0-11=0. \nonumber \end{equation}
The solutions to this quartic equation are all real, namely,
\begin{equation}  \cos \theta_0 = \frac{1\pm 3\sqrt{5}}{8} \quad {\rm and} \quad \cos \theta_0 = \frac{1}{2} \nonumber \end{equation}
the last value being a double root.  The correct cosine value for the first point of the B-map counter-clockwise 3-cycle is $\cos\theta_0=\frac{1+3\sqrt{5}}{8}$.  The other solutions are extraneous, since they don't satisfy the original equation before the squaring operation was performed.  The corresponding $\theta_0$ value is about $15.52^\circ$.

The Cartesian coordinates of the initial Kasner solution $(\Sigma_+,\Sigma_- )=(\cos\theta_0,\sin\theta_0)$ can then be computed easily since
\begin{equation}
\sin\theta_0  =  \textstyle\sqrt{1-(\frac{1+3\sqrt{5}}{8})^2 } = \textstyle\frac{\sqrt{3}}{8}(\sqrt{5}-1). \nonumber \end{equation}

By using $\cos\theta_1=\cos⁡(\theta_0+\frac{2\pi}{3})$ and similar relations, the coordinates of all the points of the 3-cycle can be found. 
 The results are:
\begin{eqnarray}
 & & (\Sigma_{+0},\Sigma_{-0})  =  \textstyle  (\cos\theta_0,\sin\theta_0)=(\frac{1+3\sqrt{5}}{8},\frac{\sqrt{3}(\sqrt{5}-1)}{8}) \nonumber \\
 & & \label{cycleSigmapts}(\Sigma_{+1},\Sigma_{-1}) = \textstyle  (\cos\theta_1,\sin\theta_1)=(\frac{1-3\sqrt{5}}{8},\frac{\sqrt{3}(\sqrt{5}+1)}{8}) \\
 & & (\Sigma_{+2},\Sigma_{-2}) = \textstyle  (\cos\theta_2,\sin\theta_2)=(-\frac{1}{4},-\frac{\sqrt{3}\sqrt{5}}{4}). \nonumber
\end{eqnarray}
The clockwise 3-cycle angles can be computed similarly or found simply by recognizing from Figure \ref{Bmap3cycle} that these angles are obtained as the mirror image of the counter-clockwise angles
reflected across the $\Sigma_+$-axis.

\subsection{The 3-Cycle: Numerical Simulations}\label{3-cycle_into_Einstein}

\begin{figure}
\begin{center}
\vspace{-0.1cm}
\includegraphics[width=7.7cm]{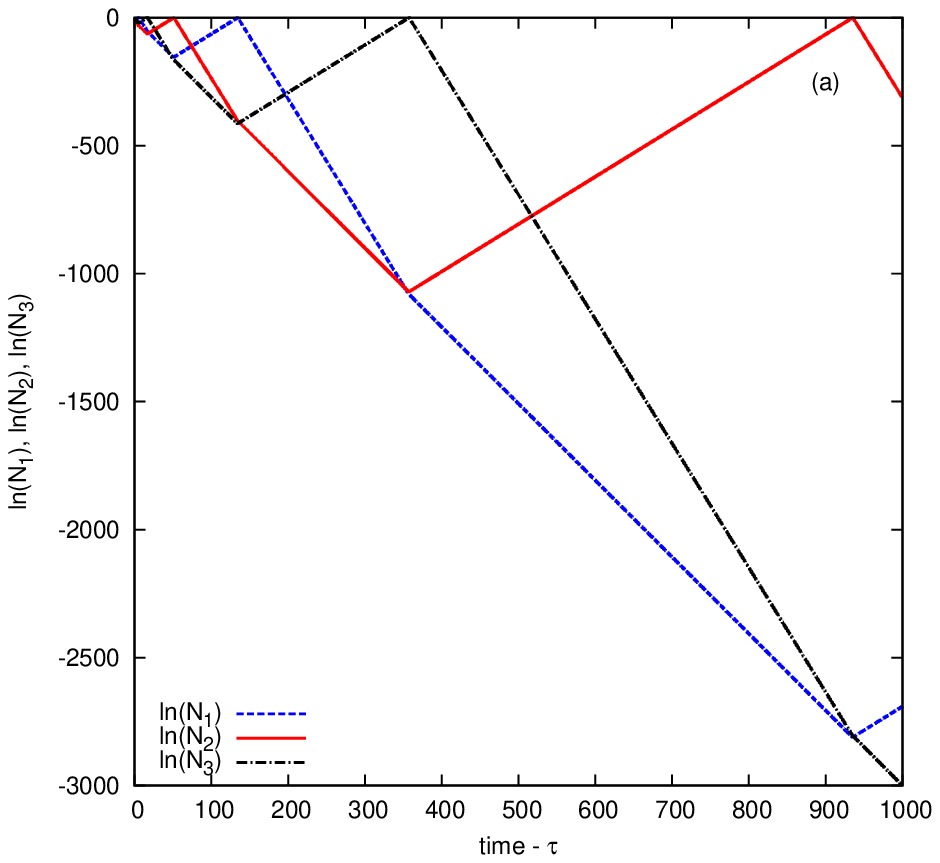}
\includegraphics[width=7.7cm]{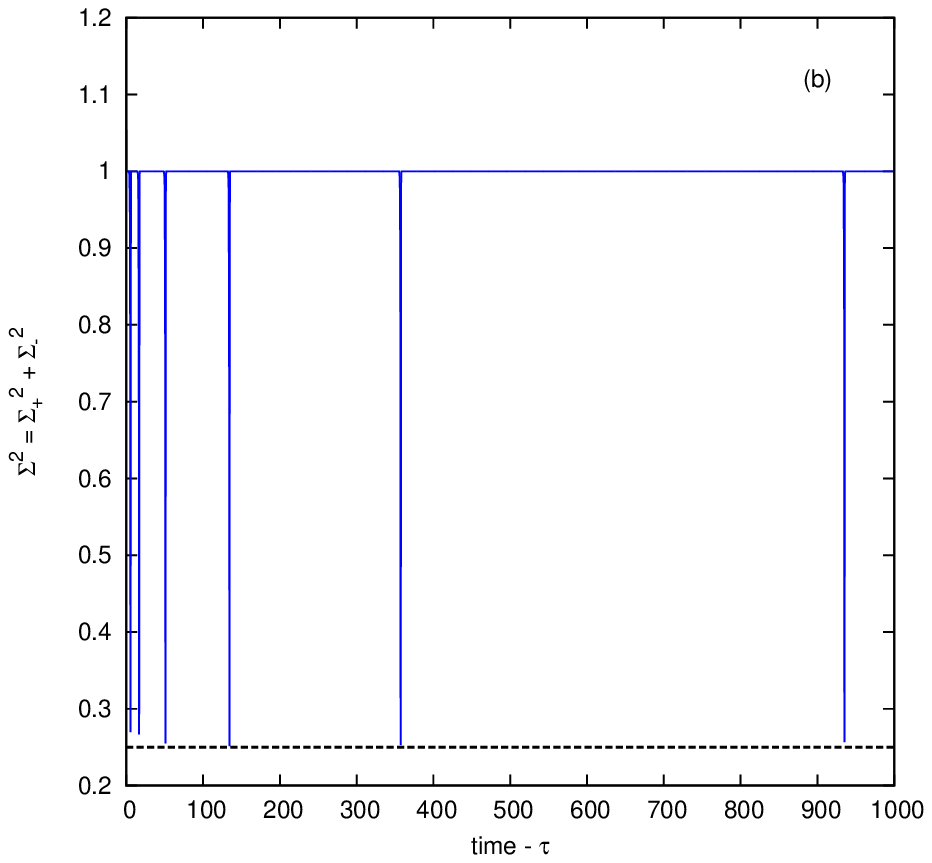}
\vspace{+0.0cm}
\caption{Numerical solutions for the curvature and shear for initial conditions close to the counter-clockwise 3-cycle solution. (a) The time series for $\ln(N_i)$.  (b) The time series for $\Sigma^2=\Sigma_+^2+ \Sigma_-^2$.} \label{threecycleZi}
\vspace{-0.5cm}
\end{center}
\end{figure}

Initial conditions for numerical solutions to the full dynamics that produce the B-map 3-cycles are rather special. If they are close to the
Kasner solution then the values of $\Sigma+$ and $\Sigma_-$ are given by one of the pairs in (\ref{cycleSigmapts}). A discussion of the choice for 
initial spatial curvature values is provided in Appendix A. However when two $N_is$ are close to zero and another has a value between zero
and one, a reasonably close approximation to the 3-cycle dynamics of all five variables can be obtained. It can be expected that the simulations will show some
sign of repetition and that this is the case is shown in Figures \ref{threecycleZi}(a) and (b) where a repetitive pattern of self-similar rectangles occurs
in the $\ln (N_i)$ time series.
Another characteristic of this graph is that the oscillations take place only in the triangular upper-half of the plot.  
In addition, the time-series of $\Sigma^2$ indicates that the value 0.25 acts as a lower bound on the dip in $\Sigma^2$ 
that takes place when the epoch (and era) transition takes place. 
In what follows it will be  
shown that not only are these observations exactly true, but that the rectangles shown in Figure \ref{threecycleZi}(a) are self-similar golden rectangles.

\subsection{The 3-Cycle: The time derivatives of the $Z_i$s}\label{FindZiPrimes}

In order to prove that the self-similar rectangles appearing in the numerical simulations are golden rectangles it is necessary to know the asymptotic values of
the time derivatives of the $Z_i$s. These can be computed directly from the ODE's using the fact that close to the singularity $q=2$, and the 3-cycle coordinates 
for $\Sigma_\pm$ are given in (\ref{cycleSigmapts}) for each $\theta$.

For $\theta_0$:
\begin{eqnarray*}
 Z_1' & = & \textstyle (2)-4(\frac{1+3\sqrt{5}}{8})=3(\frac{1-\sqrt{5}}{2})=3(-\frac{1}{\phi}) \\
 Z_2' & = & \textstyle(2)+2(\frac{1+3\sqrt{5}}{8})+2\sqrt{3}(\frac{\sqrt{3}(\sqrt{5}-1)}{8})=3(\frac{1+\sqrt{5}}{2})=3(\phi) \\
 Z_3' & = & \textstyle (2)+2(\frac{1+3\sqrt{5}}{8})-2\sqrt{3}(\frac{\sqrt{3}(\sqrt{5}-1)}{8})=3(1),
\end{eqnarray*}
for $\theta_1$,
\begin{eqnarray*}
 Z_1' & = & \textstyle (2)-4(\frac{1-3\sqrt{5}}{8})=3(\phi) \\
 Z_2' & = & \textstyle(2)+2(\frac{1-3\sqrt{5}}{8})+2\sqrt{3}(\frac{\sqrt{3}(\sqrt{5}+1)}{8})=3(1) \\
 Z_3' & = & \textstyle (2)+2(\frac{1-3\sqrt{5}}{8})-2\sqrt{3}(\frac{\sqrt{3}(\sqrt{5}+1)}{8})=3(-\frac{1}{\phi}),
\end{eqnarray*}
and for $\theta_2$,
\begin{eqnarray*}
 Z_1' & = & \textstyle (2)-4(-\frac{1}{4})=3(1) \\
 Z_2' & = & \textstyle(2)+2(-\frac{1}{4})+2\sqrt{3}(-\frac{\sqrt{3}\sqrt{5}}{4})=3(-\frac{1}{\phi}) \\
 Z_3' & = & \textstyle (2)+2(-\frac{1}{4})-2\sqrt{3}(-\frac{\sqrt{3}\sqrt{5}}{4})=3(\phi).
\end{eqnarray*}

These results are shown in Table \ref{ThreeCycleSlopes} along with similar ones obtained for the clockwise cycles obtained by changing the signs of the
$\Sigma_-$ values but keeping the signs of the $\Sigma_+$ coordinates fixed. It is clear that the slopes for the clockwise cycle are the same as in the counter-clockwise cycle, except that the indices 2 and 3 are exchanged. 

Given the Gauss map connection to the golden ratio its appearance in the EMW approach is not totally unexpected. The plots shown in Figure \ref{threecycleZi}(a)
are for the $\ln(N_i)$s and since $Z_i = -\ln{N_i}$ the signs of the slopes given in Table \ref{ThreeCycleSlopes} are opposite those appearing in the time series plots. 
The increasing slopes in shown in Figure \ref{threecycleZi}(a)
 are ``shallowest'' with a value of $3\phi^{-1}$.  The steepest decreasing slopes 
have values $-3\phi$ and a transition
occurs when these intersect with the other deceasing slope with a value of exactly $-3$.  It is a line with this latter slope that acts as the lower bound on all
the 3-cycle oscillations.
That overlaying the three curves produce golden rectangles with slopes that are related to the golden ratio, its inverse and the product of the two
is quite remarkable. 

\begin{table}
\caption{Time derivatives of $Z_i$s (wrt $\tau$) for angles representing 3-cycle points on the Kasner Ring.  Values are given for both the counter-clockwise (ccw) and clockwise (cw) cycles.  Note that the two cycles differ only by an exchange of row order, reflective of the order in which the triangle vertices of the B-map are encountered.}

\begin{center}
\begin{tabular}{crrrrrrrr}
 \hline\hline
&& \multicolumn{3}{c}{CCW} & & \multicolumn{3}{c}{CW} \\ \cline{3-5}\cline{7-9}
&& $\theta_0$ & $\theta_1$ & $\theta_2$ && $\theta_0'$  & $\theta_1'$ & $\theta_2'$  \\ \hline
$Z_1'$ && $3(-\frac{1}{\phi})$ & $3(\phi)$ & $3(1)$ && $3(-\frac{1}{\phi})$ & $3(\phi)$ & $3(1)$  \\ 
$Z_2'$ && $3(\phi)$ & $3(1)$ & $3(-\frac{1}{\phi})$ && $3(1)$ & $3(-\frac{1}{\phi})$ & $3(\phi)$ \\ 
$Z_3'$ && $3(1)$ & $3(-\frac{1}{\phi})$ & $3(\phi)$ && $3(\phi)$ & $3(1)$ & $3(-\frac{1}{\phi})$ \\ \hline\hline
\end{tabular}
\vspace{-0.2cm}
\label{ThreeCycleSlopes}
\end{center}
\end{table}

\section{The Self-Similar Golden Rectangle Structure}\label{AnalyticConfirm}

\begin{figure}
\begin{center}
\vspace{-0.4cm}\includegraphics[width=10.0cm]{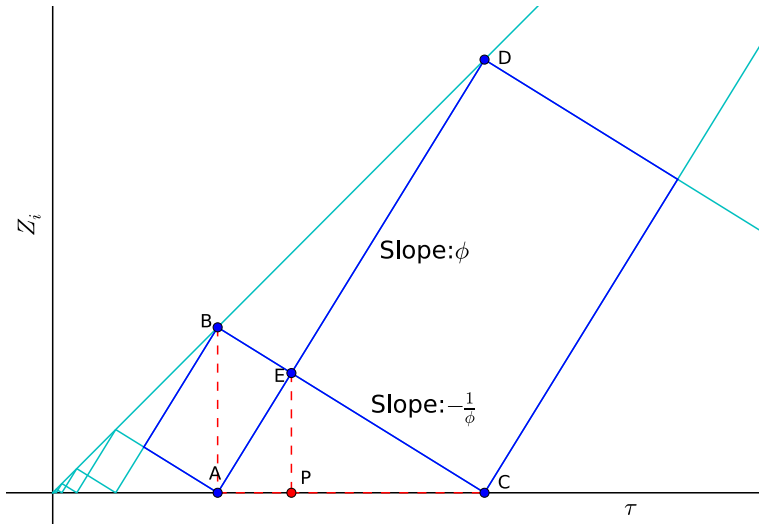}
\vspace{-0.4cm}
\caption{Two rectangles of the rescaled, reflected time series pattern with specially labelled points. The line 
connecting points $B$ and $D$ has unit slope.}  \label{TwoRect}
\vspace{-0.4cm}
\end{center}\end{figure}

In Section \ref{FindZiPrimes}, it was discovered that the time series' formed by the $Z_i$s have slopes related to the golden ratio.  
In this section, the patterns appearing in Figure \ref{threecycleZi} are explored more thoroughly, in order to prove that the rectangles
appearing in the time series' form a sequence of discrete self-similar golden rectangles.  For simplicity, the timescale, $\tau$, is rescaled to remove the factor of 
three that appears in the table. This is equivalent to scaling the overall Hubble expansion by the volume rather the average ``radius'' of the universe.
The normalization presented in \cite{WE} leads to the same result.
In addition the plots shown in Figure \ref{TwoRect} time series of the $Z_i$s so that the sign of the slopes is consistent with those given in
Table \ref{ThreeCycleSlopes}.
Figure \ref{TwoRect} will act as a reference in what follows and attention will be focused on the two largest
rectangles shown in that figure. 

The following propositions summarize some observations regarding the structure of the time series' of the curvature and shear variables
presented above. Proofs of these propositions will be provided sequentially and require only some simple geometric arguments based on the
properties of the B-map, as well as the original ODE's. 
\medskip

\noindent {\bf Proposition 1}
 The $N_i(\tau)$ variables follow straight-line segments that form a sequence of quadrilaterals.  The highest and lowest vertices of each quadrilateral are
aligned (i.e.~they have the same $\tau$-coordinate).  The slopes of the line segments forming the quadrilaterals, repeatedly cycle through 
the sequence: $1$, $-\frac{1}{\phi}$, $\phi$ ($\tau$ rescaling has removed the factor of 3).  The quadrilaterals are rectangles.
\medskip

\noindent {\bf Proposition 2}
 The highest vertex of each rectangle, lies on the line $Z_i=\tau$.
\medskip

\noindent {\bf Proposition 3}
 The lowest vertex of each rectangle, lies on the line $Z_i=0$.  (In addition the minimum value of $\Sigma^2$ during a 3-cycle transition is exactly $\frac{1}{4}$, verifying the observation made at the end of section \ref{3-cycle_into_Einstein}.)
\medskip

\noindent {\bf Proposition 4}
 Each rectangle of the sequence is golden (i.e~the length-to-width ratio of the rectangle is $\phi$).
\medskip

\noindent {\bf Proposition 5}
 The rectangles of the sequence are self-similar with a linear scaling factor of $\phi^2$. 
\medskip

\subsection{Proof of Proposition 1}

Each Kasner epoch is described by $\Sigma^2=1$, $N_i=0$ and these were the conditions applied in Section \ref{FindZiPrimes} to find the
time derivatives of the $Z_i$s.  Each Kasner epoch therefore has a constant ${\rm d} Z_i/{\rm d} \tau$ which makes the plots of $\ln(N_i)s$ straight lines.
The transitions take place simultaneously and the vertices of the quadrilaterals which are formed by the epoch transitions therefore 
occur at the same time, $\tau$.  Thus line segments $\overline{AB}$ and $\overline{CD}$ are both perpendicular to the $\tau$-axis.  

It was also proven that the time derivatives ${\rm d} Z_i/{\rm d} \tau$ cycle through the values $1$, $-\frac{1}{\phi}$, $\phi$. From the 
figure it is clear that all vertices $A$, $B$, $C$ and $D$ as well as the intersection point $E$, are points where two line segments meet: one
with positive slope $\phi$ and one with negative slope $-1/\phi$.  Therefore at each of these points the product of the two slopes is
$\phi \times (-\phi^{-1}) = -1$, i.e.~the line segments meet at right angles, making all the quadrilaterals rectangles.

\subsection{Proof of Proposition 2}
The line segments on the graph that are not a part of the rectangle sequence, also take part in the Kasner transitions and therefore must 
intersect the rectangle vertices.   The slope of these line segments must all be unity and connect to each other to form a single
line $Z_i = \tau$. Since the $\tau = 0$ point is arbitrarily chosen, a time translation can always be performed to re-set the initial time
to ensure that the intercept with the $Z_i$ axis occurs at $Z_i = 0$.

Strictly speaking Proposition 2 is not true for all time since the choice of initial values at $\tau = 0$ occurs at some finite time after the
big bang singularity and the dynamics at that time might not be consistent with the B-map description.  
However only the conditions near the singularity are being considered so that $\tau=0$ can be chosen to be at a time that is reasonably close to the big bang.  In this case, the higher vertex of each rectangle will lie on the line $Z_i=\tau$, so statement 2 holds asymptotically. 

\subsection{Proof of Proposition 3}

Proposition  3 asserts that the lower of the two vertices of each rectangle lies on the line $Z_i=0$.  This can be proven using (\ref{ellipsoideqn}), which governs the behaviour at the time of the transition between epochs.  Note that the variable $\tau$ is not involved in the B-map and the rescaled $\tau$ is irrelevant for this calculation. All that is required is a proof that $Z_i$ cannot be negative (or equivalently $N_i \leq 1$ 
since $Z_i = - \ln(N_i)$.)

\begin{figure}
\begin{center}
\vspace{-0.4cm}
\includegraphics[width=12.0cm]{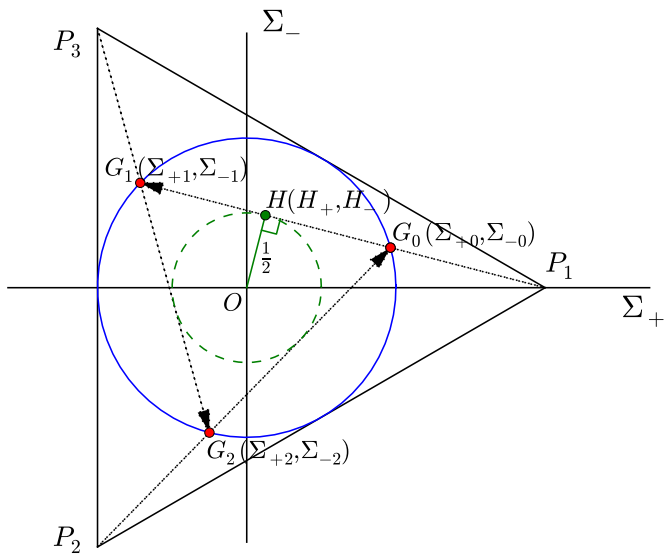}
\vspace{-0.4cm}
\caption{The diagram of the counter-clockwise 3-cycle from Figure \ref{Bmap3cycle}(a), but enhanced for use in proving Proposition 3. } \label{MinRadius}
\vspace{-0.4cm}
\end{center}
\end{figure}

The points on the Kasner ring corresponding to the counter-clockwise 3-cycle are labelled by their angular coordinates $\theta_0$, $\theta_1$, and $\theta_2$, as shown in Figure \ref{Bmap3cycle}a.  In this section, the points are renamed as $G_0$, $G_1$, and $G_2$, which can be given in their
Cartesian representation. Their coordinates are:
\begin{eqnarray}
 & & G_0 = \textstyle (\Sigma_{+0},\Sigma_{-0})=(\frac{1+3\sqrt{5}}{8},\frac{\sqrt{3}(\sqrt{5}-1)}{8}) \nonumber \\ 
 & & G_1 = \textstyle  \label{cycleSigmaptscopy}(\Sigma_{+1},\Sigma_{-1})=(\frac{1-3\sqrt{5}}{8},\frac{\sqrt{3}(\sqrt{5}+1)}{8}) \\
 & & G_2 = \textstyle  (\Sigma_{+2},\Sigma_{-2})=(-\frac{1}{4},-\frac{\sqrt{3}\sqrt{5}}{4}). \nonumber
\end{eqnarray}
See Figure \ref{MinRadius} for a version of Figure \ref{Bmap3cycle}(a) with the name changes.  

Consider the transition from $G_0$ to $G_1$ which has $P_1$ as the projection point, and consequently has a non-zero $N_1$ and $N_2=N_3=0$ for the
Taub solution describing the full dynamics of the transition.  The 3D path followed by the curve, travels along the surface of the ellipsoid given by (\ref{ellipsoideqn}), where $i=1$.

Proving statement 3 is a matter of finding how small $Z_1$ can get, (specifically, it must be proved that zero is the minimum $Z_1$ value, and that it is attained).  From (\ref{Zidefn}), $Z_1=-\ln(N_1 )$, so proving $Z_1\ge0$ is equivalent to proving $N_1\le1$.

The Bianchi IX cosmologies require that all the curvature variables have the same sign \cite{WE} (here we chose them to 
be positive) therefore (\ref{ellipsoideqn}) implies $N_1=\sqrt{\frac{4}{3}(1-\Sigma^2)}$.  So, $N_1\le1$ just requires
$ \textstyle\sqrt{\frac{4}{3}(1-\Sigma^2)}  \le  1 $ 
or 
\begin{equation}  \Sigma^2  \ge  \frac{1}{4}. \nonumber \end{equation}

Therefore, it is sufficient to prove that $\Sigma^2\ge\frac{1}{4}$ during a 3-cycle Kasner transition and that $\Sigma^2 = \frac{1}{4}$ is attained
during each transition.
$\Sigma^2$ is the square of the distance of a point from the origin of the plane, so it must be proved that the minimum distance between the line through $G_0$ and $G_1$ and the origin equals $\frac{1}{2}$, (see Figure \ref{MinRadius}).

Referring to Figure \ref{MinRadius}, by the chord-radius theorem in circle geometry, $\overline{OH}\bot\overline{G_0G_1}$ implies  $\overline{OH}$ bisects $\overline{G_0G_1}$.  In other words, $H$ can be calculated as the midpoint of $\overline{G_0G_1}$ and its magnitude taken, yielding the minimum distance.
\begin{eqnarray*}
 H & = & (H_+,H_-)\\
 & = & \textstyle(\frac{1}{2}[\Sigma_{+0} + \Sigma_{+1}],\frac{1}{2}[\Sigma_{-0}+\Sigma_{-1}]) \\
 & = & \left(\frac{1}{8},\frac{\sqrt{15}}{8}\right)
\end{eqnarray*}
Computing the radius of a circle containing this point leads to
$$ |H|  =  \sqrt{(H_+)^2+(H_-)^2} = \frac{1}{2} $$
This proves the result for $Z_1$ and shows that the min($\Sigma^2) = H^2 = \frac{1}{4}$ exactly. 

The minimum of zero for $Z_2$ and $Z_3$ (occurring during projections from $P_2$ and $P_3$, respectively), also holds, by symmetry since a rotation of the B-map diagram by $\frac{2\pi}{3}$ or $\frac{4\pi}{3}$ leaves the geometry unaltered. 

\subsection{Proof of Proposition 4}

Referring to Figure \ref{TwoRect}, two arbitrary consecutive rectangles of the sequence are shown, with labelled points.  Using a similar triangle argument,
\begin{equation}
\left. \begin{array} {l} \overline{AB}\| \overline{EP} \Rightarrow \angle BAE = \angle PEA  \\  \angle BEA = \angle APE = 90^\circ \end{array}\right\} \Rightarrow\bigtriangleup BEA\sim\bigtriangleup APE
\label{ProveSimilar1}
\end{equation}
\begin{equation}
\therefore\frac{\overline{AE}}{\overline{BE}}=\frac{\overline{EP}}{\overline{AP}}=\phi\qquad\mbox{(since $\overline{AE}$ has slope $\phi$), }
\label{ProveGoldenRect}
\end{equation}
proving that the smaller rectangle (which was arbitrarily chosen), has a length-to-width ratio of $\phi$, and is therefore golden.  Therefore, all rectangles in the sequence are golden, as required.

\subsection{Proof of Proposition 5}

Once again, using a similar triangle argument and referring to Figure \ref{TwoRect},
\begin{equation}
\hskip-1.5cm
\left. \begin{array} {l}\bigtriangleup BEA\sim\bigtriangleup APE \quad \mbox{(by (\ref{ProveSimilar1}))} \Rightarrow \angle ABE = \angle EAC  \\  \angle BEA = \angle AEC = 90^\circ \end{array}\right\} \Rightarrow\bigtriangleup BEA\sim\bigtriangleup AEC
\label{ProveSimilar2}
\end{equation}
\begin{equation}
\therefore\frac{\overline{EC}}{\overline{BE}}=\frac{\overline{EC}}{\overline{AE}}\frac{\overline{AE}}{\overline{BE}}=\phi\centerdot\phi\quad \mbox{(by (\ref{ProveGoldenRect}) and (\ref{ProveSimilar2}))} = \phi^2
\label{ProveSeqScaleup}
\end{equation}
proving (since these are two arbitrary consecutive rectangles of the sequence), that rectangles in the sequence scale (in their linear dimensions), one to the next, by a factor of $\phi^2$. 

\section{Discussion}

Using modified Ellis-MacCallum-Wainwright variables, this paper considered a Bianchi IX cosmology near its initial singularity.  The B-map, an iterative map with interesting geometric structure, is defined in this context, and was used to find the shear variable values for the 3-cycle.  These values when used as initial conditions for the
 Einstein equations (the full set of ODEs) produced a time series representation that exhibited a discrete self-similar golden rectangle structure. 
In addition it was found directly from the ODE's that the time derivatives of the logarithms of the spatial curvature variables always equalled $-\phi$, $\phi^{-1}$ and $1$ 
in an appropriately re-scaled time variable. The geometric structure consists of a sequence of golden rectangles, where the linear scale factor from one to the next is $\phi^2$.
All of these relations are exact and can be easily proven using the asymptotic conditions on the equations governing the Bianchi IX evolution along with some simple
relations from Euclidean plane geometry.  

While the golden ratio appears in the analysis of many natural and engineered systems, often it does so in an approximate form. If there is
an exact relationship to the golden ratio it may be difficult or even impossible to discover.  That the Einstein equations lead to the
golden rato exactly is somewhat of a mystery, but
the ubiquitousness of the number $\phi$ and its appearance in a large number of totally unrelated systems remains one
of the great mysteries of mathematics.

From a mathematical point of view cosmological expansion and gravitational collapse can be considered to be related phenomena \cite{HawEll}. Discussions of discrete self-similarity in cosmological collapse have been quite rare and its relationship (if any) to self-similar collapse 
to form black holes should be better understood. For example, is there a parameter that appears in critical black hole collapse 
that is exactly the
golden ratio? Does the golden ratio appear elsewhere in the theory of general relativity?
The relationship of the BKL conjecture regarding the generic form of spacetime singularities in inhomogeneous collapse remains an open question and
hopefully the study of self-similar Bianchi-IX solutions may provide some insight into singularity formation in more general 
forms of gravitational collapse.

The golden rectangle graph structure discussed in this article is interesting in its own right. Referring to Figure \ref{TwoRect}, it can be shown that the 
construction of intersecting, self-similar rectangles that have one diagonal perpendicular to the $x$-axis such that {\bf all} the upper vertices lie on the line $y=x$,
 must be golden rectangles with a linear $\phi^2$ scaling.  
Interleaving this pattern with a similar one leads to a tiling of the Euclidean plane that can
be used to provide geometric proofs of identities involving the golden ratio, the 
Fibonacci numbers and the number $\pi$.  A manuscript describing this interesting aspect of the self-similar golden rectangle pattern
appearing the the dynamics of the vacuum Bianchi-IX cosmologies is in preparation.

\section{Acknowledgements}

Partial funding for this research was provided by an Natural Sciences and Engineering Research Council of Canada (NSERC) Discovery Grant to DH who acknowledges conversations with Adrian Burd, Teviet Creighton, Scott Macdonald, and John Wainwright concerning periodic solutions to the B-map that led to these investigations.

\appendix
\vskip 0.2in
{\Large {\bf Appendix}}

\section{Initial Conditions for Numerical Simulations} \label{NumericConfirm}

In section \ref{FindZiPrimes}, it was proven that the B-map 3-cycle $\Sigma_+$ and $\Sigma_-$ values (found in section \ref{3CyclePoints}), produce the golden-ratio-related 
time derivatives for the $Z_i$s. If one wishes to re-construct the full dynamics of the 3-cycle then the $\Sigma_\pm$ values along with appropriate initial conditions
for the $Z_i$s are required in order to obtain the correct 3-cycle evolution for a large number of transitions. Numerical errors arising from an inaccurate choice
of curvature variables eventually produce deviations away from self-similarity, typically after five to eight transitions. 
At the same time it was 
found that the minimum value of $\Sigma^2$ would dip below the value of $\frac{1}{4}$.  

\begin{figure}
\begin{center}
\vspace{-0.4cm}
\includegraphics[width=12.0cm]{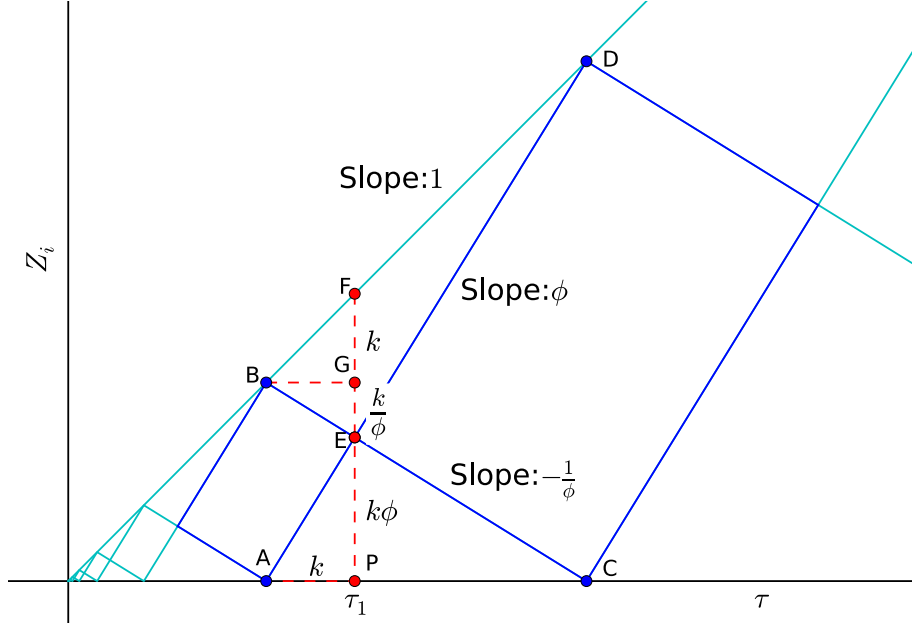}
\vspace{-0.4cm}
\caption{Showing how initial conditions in which two $Z_i$ are equal, and the third is twice their value, is consistent with the 3-cycle conditions.} \label{ZiInitRatio}
\vspace{-0.4cm}
\end{center}
\end{figure}

What is required to obtain results that can be sustained over a longer number of transitions is to begin with the best possible set of initial conditions that are
consistent with the B-map 3-cycles and the transition values of the $N_i$s. A bit of hindsight based upon what is known about the self-similar structure helps
in choosing the appropriate initial conditions.
Figure \ref{ZiInitRatio} shows a section of the self-similar golden rectangle sequence with some points labelled and the slope values shown.  We set $\tau=\tau_1$ 
as the point in time during an epoch at which two of the $Z_i$ values are equal.  If $k$ is the time since the last transition, then the vertical line  segments at time 
$\tau_1$ are as shown. First $\Delta BGF$ is a 45-45-90 triangle therefore $\overline{BG} = \overline{FG} = k$. The right triangles $\Delta AEP$ and $\Delta EBG$ 
are both similar to $\Delta BAE$ which has two perpendicular sides obeying
$\overline{AE}/\overline{BE} = \phi$.  This makes $\overline{EP} = k\phi$ and $\overline{GE} = k/\phi$.  
Then $\overline{FG} + \overline{GE} = k + \frac{k}{\phi} = k(1 + \frac{1}{\phi})=k\phi  = \overline{EP}$, and this shows that the 
 value of the third $Z_i$ (at $F$) is twice the common value of the other two.

Since a time translation does not affect the overall dynamics, any small value, $\tau_1$, can be taken as the initial $\tau$ value $\tau_0$.  One of the initial $Z_i$ values 
should equal $\tau_0$, and the other two should equal $\tau_0 / 2$.  For a given 3-cycle $(\Sigma_+,\Sigma_-)$ point there is a 
one-in-three chance of picking the correct $Z_i$ as the largest one.

For the time series shown in Figure \ref{threecycleZi} the choice made was $\tau_0=0.1$; initially $(\Sigma_+,\Sigma_-)=(\Sigma_{+2},\Sigma_{-2})$ (see (\ref{cycleSigmapts})); and $Z_1=\tau_0 /2$, $Z_2=\tau_0$ and $Z_3=\tau_0 /2$.  Even after ten transitions between $\tau = 0.1$ and $\tau = 1000.$, there was no significant deviation away from
the discrete self-similar pattern and the minimum value for $\Sigma^2$ never fell below one-quarter.  


\end{document}